\documentclass[12pt]{JHEP3}
\usepackage{graphics,latexsym,amsmath,epsfig,bm}

\newcommand{\as}{\alpha_s}
\newcommand{\bas}{\bar{\alpha}_s}
\newcommand{\e}{\mathrm{e}}
\newcommand{\shat}{\hat{s}}
\newcommand{\ep}{\epsilon}

\renewcommand{\d}{\mathrm{d}}
\newcommand{\A}{\mathcal{A}}
\newcommand{\B}{\mathcal{B}}
\newcommand{\M}{\mathcal{M}}
\newcommand{\F}{\mathcal{F}}
\newcommand{\G}{\mathcal{G}}

\renewcommand{\O}{\mathcal{O}}
\newcommand{\C}{\mathbf{C}}
\newcommand{\ka}{\mathbf{k}}
\newcommand{\q}{\mathbf{q}}
\newcommand{\el}{\bm{\ell}}
\newcommand{\Tr}{\mathrm{Tr}}
\newcommand{\pDL}{$\pi^2 DLL$ }
\newcommand{\pDLA}{$\pi^2 DLLA$ }

\renewcommand{\Re}{\mbox{$\mathrm{I\!Re}$}}
\renewcommand{\Im}{\mbox{$\mathrm{I\!Im}$}}
\newcommand{\Cross}{{\,\scalebox{0.67}{\rotatebox{90}{$\Join$}}}}

\renewcommand{\vec}[1]{\underline{#1}}
\renewcommand{\S}{\mathrm{S}}
\renewcommand{\H}{\mathrm{H}}
\newcommand{\R}{\mathrm{R}}
\newcommand{\D}{\mathrm{D}}
\newcommand{\wt}{\widetilde}
\newcommand{\ev}[1]{\lambda^{(#1)}}

\title{Gaps between Jets in the High Energy Limit}
\author{J.R. Forshaw, A. Kyrieleis \\School of Physics \& Astronomy, University of Manchester, \\
  Oxford Road, Manchester M13 9PL, U.K.\\
\email{forshaw@mail.cern.ch}, \email{kyrieleis@hep.man.ac.uk}}
\author{M.H. Seymour \\School of Physics \& Astronomy, University of Manchester
\\and \\
Theoretical Physics Group (PH-TH), CERN, CH-1211 Geneva 23,
Switzerland\\
\email{Mike.Seymour@cern.ch}}

\abstract{
We use perturbative QCD to calculate the parton level cross section for the 
production of two jets that are far apart in rapidity, subject to a limitation 
on the total transverse momentum $Q_0$ in the interjet region. 
We specifically address the question of how to combine the approach which
sums all leading logarithms in $Q/Q_0$ (where $Q$ is the jet transverse momentum)
with the BFKL approach, in which leading logarithms of the scattering energy are
summed. This paper constitutes progress towards the simultaneous summation
of all important logarithms. Using an ``all orders'' matching, we are able to 
obtain results for the cross section which correctly reproduce the two approaches in 
the appropriate limits. 
}

\keywords{qcd, jet}
\preprint{CERN-PH-TH/2005-028}

\begin{document}
\section{Introduction}
Events with high-$p_T$ jets separated by large rapidity gaps at
hadron colliders are of significant theoretical interest. Studying these
events provides us with the possibility to better understand QCD in the high
energy limit and also to understand QCD radiation in ``gap'' events.
There are two major approaches to the production of two gap-separated
jets.

In the first approach, parton-parton elastic scattering with a QCD colour
singlet exchange is regarded as providing
the leading contribution to the cross-section. The 
leading-$Y$ terms ($Y$ is the rapidity interval between the jets) are summed, i.e.   
\begin{align}
\sim \as^n Y^n
\end{align} 
This is the so-called BFKL \cite{BFKL} approach to the gaps between jets process.
The observable calculated in this approach does not consider any
radiation into the interjet region. Experiments though, impose an
upper bound  on this radiation by necessity.
Calculations \cite{MueTa, CoxFo, MoMaRy, EnbInMo1, EnbInMo2} 
based on this approach have
been performed and compared with some success to the data from HERA and Tevatron
\cite{data}.

In the second approach  the energy
flow into the gap is certainly considered. In accord with experimental
observations soft radiation with
transverse energy below $Q_0$ is allowed
in the interjet region. This gives rise to  logarithms of $Q/Q_0$ where
$Q$ is the transverse momentum of the jets. The aim here is to sum all terms 
of the form
\begin{align}
\sim \as^n Y^m L^n\;,\quad m\le n
\label{LLQ terms}
\end{align}
where  $L=\ln Q^2/Q_0^2$. In particular, at each order of $\as$ the 
full $Y$-dependence of the coefficient of $L^n$ is kept.
The global leading  logarithms of $Q/Q_0$ ($LLQ_0$) have  been summed for various jet 
definitions \cite{OdSt98,Od00,BeKuSt,AbSey03} and non-global effects have been
considered in \cite{AbSey03,AbSey02}.

For asymptotically  large $Y$ at fixed $L$  the BFKL approach is expected to
be the most appropriate one. On the other hand, using a gap definition with large
values of $Q_0$ at realistic collider energies the second approach
ought to be the best. In order to get a better understanding  of
the gaps-between-jets processes  at colliders it is desirable to combine the
two approaches. This is the main issue of this paper.

To combine the two approaches order-by-order we need to
prevent double counting and make sure the divergences arising from the BFKL
approach (at each order in $\alpha_s$) cancel in the jet cross section. 
To this end we recalculate the
gap cross section with limited transverse energy-flow into the gap in
the high energy limit. In addition to the double-leading-logarithmic ($DLL$) terms
we also consider certain terms sub-leading in $Y$, namely
those arising from the imaginary parts of the loop integrals. In the first
instance, we thus sum terms of 
the type
\begin{align}
\sim\as^n L^n Y^m \pi^{n-m} \;,\quad m\le n
\end{align}
where only integer powers of $\pi^2$ appear. We term this approximation the
\pDLA. We argue that the
gap cross section can be obtained by calculating the elastic
quark-quark amplitude with all virtual gluon rapidities constrained to
the gap region and all transverse momenta constrained to be above $Q_0$. 
Based on this we
show that in the \pDLA, and also in the $LLQ_0A$, the BFKL cross
section can be included explicitly up to $\mathcal{O}(\as^5)$ and derive the
expression for the combined gap cross section. 

At this stage, we have not been able to go beyond $\mathcal{O}(\as^5)$ in fixed-order 
matching. However, we do present an all-orders cross section that smoothly interpolates 
between the $LLQ_0$ and the BFKL results. In this context, we consider three 
phenomenological matching schemes.

The remainder of the paper is structured as follows. In the next section we calculate 
the gap cross section in the \pDLA, consider some approximations to it and compare to the
$LLQ_0$ cross section. Then, we calculate the quark-quark scattering 
amplitude in the leading-$Y$ (BFKL)
approximation, deriving the general result at any fixed order in  $\as$. 
Finally we discuss the matching of the two approaches. Firstly we attempt
a fixed-order matching which we are able to accomplish to $\mathcal{O}(\as^5)$
and then we present our all-order matching results.
\section{Summing logarithms in $Q_0$}
We seek to calculate the cross section for two-jet production in the
high-energy (i.e.~high rapidity separation) limit, with limited total
scalar transverse momentum in the interjet region.  We require this
transverse momentum to be below $Q_0$, the jets to have transverse
momentum $Q$, and be separated by a rapidity interval $Y$, and consider
the region
\begin{equation}
  \Lambda_{QCD}^2 \ll Q_0^2 \ll Q^2 \ll \shat=\e^Y\,Q^2.
\end{equation}
The first of these relations is to ensure that we can use perturbation theory,
although we will ultimately be interested in extrapolating $Q_0$ towards
$\Lambda_{QCD}$, while the other two are to ensure that the logarithms
we are interested in are large. In this section,
we calculate the partonic gap cross section for the process,
$\mathrm{qq'\to qq'}$ 
\begin{equation}
  \sigma\equiv\frac{\d\sigma(\shat,Q_0,Y)}{\d Q^2}
\end{equation}
for fixed  $\shat$ in the \pDLA. Since we are not
sensitive to collinear emission, we do not need to worry about the
details of initial state collinear factorization, and only calculate
partonic cross sections. 

Our approximation implies the eikonal
 (soft gluon) approximation, valid when $Q^2\ll\shat$. 
To generate the leading logs in $Q_0$, we make the approximation
of strongly-ordered  transverse momenta for the real and virtual
gluons. The scalar sum of all gluon transverse momenta can therefore
 be approximated by the largest single transverse momentum, $\sum_i
 q_{\perp i} \approx\max_i q_{\perp i}$. We therefore require that any
 real gluon emitted in the gap region have transverse momentum below $Q_0$.

Let us denote denote by $\sigma^{(n)}$ the gap cross section at
$\mathcal{O}(\as^{n+2})$.
Clearly at zeroth order, the gap cross section coincides with the inclusive
cross section,
\begin{equation}
  \sigma^{(0)}= \frac{4\pi\as^2}{Q^4} \frac{(N_c^2-1)}{4 N_c^2}.
  \label{sig 0}
\end{equation}
As the basis for the calculation to all
orders we employ the following theorem.

\subsection{Theorem}
``At any finite order in perturbation theory, in the strongly-ordered
approximation, the gap cross section is given by the
two-to-two cross section, with all virtual gluon rapidities constrained
to the gap region and transverse momenta above $Q_0$, and with the
exchanged gluons attached only to the external lines.''\\

This is clearly a major simplification, since it means we never have to
calculate any real emission or triple-gluon-vertex diagrams.

Central to the proof of this theorem is the following result. For every cut
through a given cut diagram, and every way of attaching a softer gluon
to the external lines of that diagram, we can always make four cuts: in
which the soft gluon is to the left of the cut; to the right of the cut;
crossing the cut from left to right; and crossing the cut from right to
left.  The contributions from the real parts of the loop integral are
equal and opposite to the real emission integral.  The imaginary part of the loop
integral, if present, cancels between the cuts to the left and right of
the gluon.  The total contributions are therefore equal and opposite.

It is then straightforward to see that we only get a non-zero
contribution from configurations in which all gluons, real or virtual,
have transverse momentum above $Q_0$.  If the softest gluon is below
$Q_0$, then the value of the observable is the same, whether it is real
or virtual.  The cancellation is therefore able to take place and we get
zero contribution.  Therefore the softest gluon must be above $Q_0$,
therefore all gluons must be above $Q_0$.

We can also use the same argument to prove that the softest gluon must
be within the gap region.  The same argument does not however apply to
harder gluons, as it is not guaranteed that for them all four cuts
listed above are consistent with strong ordering. 
We assume here, as in conventional calculations of
gaps-between-jets processes, that contributions from outside the gap region
cancel, but note that our recent work has cast doubt on this assumption \cite{inprep}.
Therefore, all gluons must be within the gap  region. 
Finally, since for a real gluon in the gap region our observable is
zero, all gluons must be virtual.

\subsection{Calculation in the \pDLA}

We therefore have to calculate the all-orders amplitude for quark
$2\to2$ scattering, with the phase space for the gluons constrained to
the gap region in rapidity and with transverse momentum above~$Q_0$.
The loop integrals themselves are trivial~-- they are just nested sets
of integrals.  The
only complication concerns the colour structure.

Let us denote by $\A\C(Q_0)$ the amplitude defined above, including its
colour factor.  It can be decomposed into singlet and octet components,
\begin{equation}
  \A\C(Q_0) = \A_1(Q_0)\,\C_1+\A_8(Q_0)\,\C_8.
\end{equation}
The gap cross section we are interested in is therefore proportional to
\begin{equation}
  \sigma\propto
  |\A_1(Q_0)|^2\Tr\{\C_1^\dagger\C_1\}+|\A_8(Q_0)|^2\Tr\{\C_8^\dagger\C_8\}
  =|\A_1(Q_0)|^2N_c^2+|\A_8(Q_0)|^2\frac14(N_c^2-1).
  \label{propto}
\end{equation}
We denote the $n$th order contribution to the colour amplitudes as
$\A_i^{(n)}$.  Since for $Q_0=Q$ the transverse momentum integrals give
zero, we must have
\begin{equation}
  \A_i(Q)=\A_i^{(0)}.
\end{equation}
Since the zeroth order contributions $\A_i^{(0)}$ are given by single
gluon exchange, we have $\A_1^{(0)}=0$.  We choose to normalize
$\A_8^{(0)}$ in such a way that the constant of proportionality in
(\ref{propto}) is unity, i.e.
\begin{equation}
  \A_8^{(0)}=\sqrt{\frac{\sigma^{(0)}}{\frac14(N_c^2-1)}}.\label{A8 norm}
\end{equation}

We introduce a simple vector notation, with basis vectors $\vec e_1=
\C_1$ and $\vec e_2=\C_8$, such that
\begin{equation}
  \vec\A=\left(\begin{array}{c}\A_1\\\A_8\end{array}\right),
\end{equation}
and we can write
\begin{equation}
  \sigma = \vec\A^\dagger(Q_0)\,\S\,\vec\A(Q_0),
\end{equation}
with
\begin{equation}
  \S =
  \left(\begin{array}{cc}N_c^2&0\\0&\frac14(N_c^2-1)\end{array}\right).
\end{equation}

We explicitly calculate the amplitude at $\mathcal{O}(\as^2)$ and
$\mathcal{O}(\as^3)$. Note that, besides the (real) leading-$Y$ parts of the
amplitudes we also keep the sub-leading terms which differ from the
leading parts in the powers of $Y$ and $\pi$, see appendix. As
for the one-loop amplitude $\vec\A^{(1)}(Q_0)$ the soft gluon exchange changes the
colour of the external quarks, so that the one-loop correction can be
considered to be a matrix acting on $\vec\A^{(0)}$ to obtain
$\vec\A^{(1)}$. Similarly, $\vec\A^{(2)}$ can be obtained from $\vec\A^{(1)}$.  We can
therefore write the result in matrix form as
\begin{equation}
  \vec\A^{(1)}(Q_0) = -\frac{2\as}{\pi}
  \int_{Q_0}^Q\frac{\d k_\perp}{k_\perp} \Gamma\,\vec\A^{(0)},
\end{equation}
and
\begin{eqnarray}
  \vec\A^{(2)}(Q_0) &=& \left(\frac{2\as}{\pi}\right)^2
  \int_{Q_0}^Q\frac{\d k_{\perp1}}{k_{\perp1}}
  \int_{k_{\perp1}}^Q\frac{\d k_{\perp2}}{k_{\perp2}}
  \Gamma^2\,\vec\A^{(0)}\nonumber
\\&=& \frac1{2!}\left(-\frac{2\as}{\pi}
  \int_{Q_0}^Q\frac{\d k_\perp}{k_\perp}\Gamma\right)^2
  \vec\A^{(0)}
\end{eqnarray}
with
\begin{equation}
  \Gamma = \left(\begin{array}{cc}\quad 0\quad&\frac{N_c^2-1}{4N_c^2}i\pi\\
    \quad i\pi\quad&\frac{N_c}2Y-\frac1{N_c}i\pi\end{array}\right).
\end{equation}
This can be summed to all orders by inspection, to give
\begin{equation}
  \vec\A(Q_0) = \exp\left\{-\frac{2\as}{\pi}
  \int_{Q_0}^Q\frac{\d k_\perp}{k_\perp} \Gamma\right\}\vec\A(Q),
\label{result ampl}
\end{equation}
where the exponential of a matrix is defined by its power series.
 This result can also be derived by considering the evolution
equation
\begin{equation}
  \vec\A(Q_0) = \vec\A(Q) -
  \frac{2\as}{\pi}
  \int_{Q_0}^Q\frac{\d k_\perp}{k_\perp}\Gamma\vec\A(k_\perp),
\end{equation}
which can easily be shown to be solved by the expression above.

We therefore have
\begin{equation}
  \sigma=\vec\A^\dagger(Q)\exp\left\{-\frac{2\as}{\pi}
  \int_{Q_0}^Q\frac{\d k_\perp}{k_\perp} \Gamma^\dagger\right\}
  \S\exp\left\{-\frac{2\as}{\pi}
  \int_{Q_0}^Q\frac{\d k_\perp}{k_\perp} \Gamma\right\}\vec\A(Q).
\end{equation}
As a final step, we note that a scalar can be written as its own trace,
then the cyclicity of the trace can be used to bring $\A$ to the
beginning, and finally, we can define
\begin{equation}
  \vec\A(Q)\vec\A^\dagger(Q) \equiv \H = \left(\begin{array}{cc}0&0\\0&
    \frac{\sigma^{(0)}}{\frac14(N_c^2-1)}\end{array}\right),
\label{H}
\end{equation}
to give
\begin{equation}
  \sigma=\Tr\left[\H\exp\left\{-\frac{2\as}{\pi}
  \int_{Q_0}^Q\frac{\d k_\perp}{k_\perp} \Gamma^\dagger\right\}
  \S\exp\left\{-\frac{2\as}{\pi}
  \int_{Q_0}^Q\frac{\d k_\perp}{k_\perp} \Gamma\right\}\right].
\label{structure}
\end{equation}

If we neglect the imaginary parts of the anomalous dimension matrix, it
becomes diagonal, so that the matrix structure collapses and the result
is trivial,
\begin{equation}
  \sigma\stackrel{i\pi\to0}\longrightarrow
  \sigma^{(0)}\exp\left\{-N_c\frac{\as}{\pi}LY\right\} \equiv \sigma_{DL},
\label{no pi}
\end{equation}
where $L=\ln\frac{Q^2}{Q_0^2}$. (\ref{no pi}) is the gap cross section
in the DLLA since the neglected imaginary parts of $\Gamma$ are
the terms sub-leading in $Y$. 

However, we are interested in the solution with the imaginary
parts.  This can be obtained by diagonalizing the anomalous dimension
matrix $\Gamma$, since then the exponentials become straightforward.  We
define $\R$ to be the matrix that diagonalizes $\Gamma$,
\begin{equation}
  \wt\Gamma\equiv\R^{-1}\Gamma\R=
  \left(\begin{array}{cc}\ev1&0\\0&\ev2\end{array}\right),
\end{equation}
where $\ev{1,2}$ are the eigenvalues of $\Gamma$, which we order by
\begin{equation}
  \Re(\ev1) < \Re(\ev2).
\end{equation}
We can therefore insert factors of $\R\R^{-1}$ or
$\R^{-1\dagger}\R^\dagger$ at appropriate points, to obtain
\begin{eqnarray}
  \sigma&=&\Tr\left[\R^{-1}\H\R^{-1\dagger}\exp\left\{-\frac{\as}{\pi}L
  \R^\dagger\Gamma^\dagger\R^{-1\dagger}\right\}
  \R^\dagger\S\R\exp\left\{-\frac{\as}{\pi}L
  \R^{-1}\Gamma\R\right\}\right]
\\&=&
  \Tr\left[\wt\H\D^\dagger\wt\S\D\right],
\end{eqnarray}
where
\begin{eqnarray}
  \wt\H&=&\R^{-1}\H\R^{-1\dagger}\\
  \wt\S&=&\R^\dagger\S\R\\
  \D&=&\exp\left\{-\frac{\as}{\pi}L
  \wt\Gamma\right\}
  =\left(\begin{array}{cc}
    \exp\left\{-\frac{\as}{\pi}L
    \ev1\right\}&0\\
    0&\exp\left\{-\frac{\as}{\pi}L
    \ev2\right\}
  \end{array}\right)\label{D}.
\end{eqnarray}
For the eigenvalues, we obtain
\begin{eqnarray}
  \ev1 &=&
  \frac{N_c^3Y-2N_ci\pi-N_c^2\sqrt{N_c^2Y^2-4Yi\pi-4\pi^2}}{4N_c^2}\\ \nonumber
  &&\quad\approx\frac{N_c^2-1}{2N_c^3}\frac{\pi^2}{Y}
  +\frac{N_c^2-1}{N_c^5}\frac{i\pi^3}{Y^2}
  +\O\left(\frac1{Y^3}\right)
  ,\phantom{(99)}\\
  \ev2 &=&
  \frac{N_c^3Y-2N_ci\pi+N_c^2\sqrt{N_c^2Y^2-4Yi\pi-4\pi^2}}{4N_c^2} \\ \nonumber
&&  \approx\frac{N_c}2Y-\frac{i\pi}{N_c}
  +\O\left(\frac1{Y}\right)
  ,
\end{eqnarray}
where the approximation gives the leading real and imaginary parts in
the large-$Y$ limit.

For the diagonalization matrix, we obtain
\begin{equation}
  \R=\left(\begin{array}{cc}
    1&\frac{i\ev1}{\pi}\\
    \frac{-4iN_c^2\ev1}{(N_c^2-1)\pi}&1\\
  \end{array}\right).
\end{equation}
It is worth noting that this becomes diagonal in the large-$Y$ limit
(with leading off-diagonal terms purely imaginary and of order $1/Y$).

In terms of $\ev1$, the matrices $\wt\S$ and $\wt\H$ have fairly simple
forms,
\begin{align}
  \wt\S &= \left(\begin{array}{cc}
    N_c^2+\frac{4|\ev1|^2N_c^4}{(N_c^2-1)\pi^2}&
    \frac{2i\mathrm{I\!Re}(\ev1)N_c^2}{\pi}\\
    -\frac{2i\mathrm{I\!Re}(\ev1)N_c^2}{\pi}&
    \frac14(N_c^2-1)+\frac{|\ev1|^2N_c^2}{\pi^2}
  \end{array}\right),\label{tilde S}\\
  \wt\H &= \frac{4\sigma^{(0)}(N_c^2-1)\pi^2}{
    \left((N_c^2-1)\pi^2-4{\ev1}^2N_c^2\right)
    \left((N_c^2-1)\pi^2-4{\ev1}^{*2}N_c^2\right)}
    \left(\begin{array}{cc}|\ev1|^2&-i\pi\ev1\\
      i\pi{\ev1}^*&\pi^2
  \end{array}\right).\label{tilde H}
\end{align}
The result for the gap cross section reads
\begin{eqnarray}
  \sigma &=&\phantom{+}
  \exp\left\{-\frac{\as}\pi L
  \left(\ev1+{\ev1}^*\right)\right\}
  \wt\S_{11}\wt\H_{11}
\nonumber\\&&+
  \exp\left\{-\frac{\as}\pi L
  \left(\ev2+{\ev1}^*\right)\right\}
  \wt\S_{12}\wt\H_{21}
\nonumber\\&&+
  \exp\left\{-\frac{\as}\pi L
  \left(\ev1+{\ev2}^*\right)\right\}
  \wt\S_{21}\wt\H_{12}
\nonumber\\&&+
  \exp\left\{-\frac{\as}\pi L
  \left(\ev2+{\ev2}^*\right)\right\}
  \wt\S_{22}\wt\H_{22}.
\label{fullall}
\end{eqnarray}
Since $\wt\S$ and $\wt\H$ are Hermitian, it is easy to see that this
result is purely real, as it should be.  We do not explicitly
substitute in for $\ev1$, as the results do not particularly simplify
further.

To obtain the asymptotic large-$Y$ limit (i.e.~$Y\gg\pi$), we take the
large-$Y$ limit of the exponents and coefficients separately,
\begin{eqnarray}
  \sigma &=&\phantom{+}\sigma^{(0)}
  \exp\left\{-\frac{\as}\pi L
  \frac{N_c^2-1}{N_c^3}\,\frac{\pi^2}{Y}\right\}
  \frac{N_c^2-1}{N_c^4}\,\frac{\pi^2}{Y^2}
\nonumber\\&&-\sigma^{(0)}
  \exp\left\{-\frac{\as}{2\pi}L
  N_cY\right\}
  4\frac{N_c^2-1}{N_c^4}\,\frac{\pi^2}{Y^2}
\nonumber\\&&+\sigma^{(0)}
  \exp\left\{-\frac{\as}\pi L
  N_cY\right\}
  \left(1+3\frac{N_c^2-1}{N_c^4}\,\frac{\pi^2}{Y^2}\right).
\label{limits}
\end{eqnarray}
Clearly, the final term contains the expected exponentiation of the
lowest order term.  However, for large $Y$ or large $L$, the first
term dominates.  Indeed for large enough $Y$, the result is actually
$Q_0$- and $\as$-independent,
\begin{equation}
  \sigma \stackrel{Y\to\infty} =
  \sigma^{(0)}\frac{N_c^2-1}{N_c^4}\,\frac{\pi^2}{Y^2} \equiv \sigma_\infty.
\label{large Y}
\end{equation}
\subsection{Extending to the $LLQ_0A$}
It turns out that the extension of the previous results to the
$LLQ_0A$, i.e. including those additional terms which lie beyond the
high energy approximation, is straightforward.  The basic structure, (\ref{structure}),
remains unchanged, but one has to modify the hard and anomalous
dimension (but not the soft) matrices. The hard matrix $\H$ is sensitive to
the kinematics and dynamics of the hard scattering, for example, the
quarks' spin and flavours (whether they are an equal or unequal
flavour pair). The anomalous dimension matrix is sensitive to the hard
scattering kinematics (i.e. the scattering angle) and also to the
definition of the gap observable, in particular the shape of the
region over which the scalar transverse momentum is summed. 

In our case, since we consider a process that has only one colour flow at
lowest order ($qq'\to qq'$) the hard matrix is still proportional to
$\left(\begin{array}{cc}0&0\\0&1\end{array}\right)$ and the difference
appears only as a change in $\sigma^{(0)}$ (see \eqref{H}) which is an overall factor
that multiplies all of our results. In the following we denote  this
changed LO cross section by $\sigma_{full}^{(0)}$.
The anomalous dimension matrix, $\Gamma$, is a
function both of $\Delta y$ (the rapidity interval defining the gap) and
$\Delta\eta$ (the rapidity interval between the two jets).   In the cone
algorithm, one has
$\Delta y=\Delta\eta-2R$, where $R$ is the cone radius used to define
jets.  In this case it is natural to define $\Delta y=Y$, and obtain
\begin{equation}
  \Gamma = \left(\begin{array}{cc}
    \frac{N_c^2-1}{4N_c}\rho(Y)&\frac{N_c^2-1}{4N_c^2}i\pi\\
    i\pi&\frac{N_c}2Y-\frac1{N_c}i\pi+\frac{N_c^2-1}{4N_c}\rho(Y)
  \end{array}\right),
\end{equation}
where
\begin{align}
  \rho(Y) &= \log\frac{\sinh(\Delta\eta/2+\Delta
y/2)}{\sinh(\Delta\eta/2-\Delta y/2)} - Y\\
& = \log\frac{\sinh(Y+R)}{\sinh{R}}-Y
      \stackrel{Y\to\infty}\longrightarrow -\log(1-\e^{-2R})
      \stackrel{\mbox{\tiny$R$~large}}\longrightarrow \e^{-2R}.
\end{align}
$R=1$ is common for this kind of study (although $R=0.7$ has also been
used).  Note that $\rho(Y)$ tends to zero for small $Y$ and is
monotonic, so always remains small.
In the $LLQ_0A$ the coupling has to run. This 
can be easily incorporated into the cross sections obtained so far, simply by replacing
\begin{equation}
  \frac{\as}{\pi}L=
  \frac{2\as}\pi\int_{Q_0}^Q\frac{\d k_\perp}{k_\perp}\longrightarrow
  \frac2\pi\int_{Q_0}^Q\as(k_\perp)\frac{\d k_\perp}{k_\perp}=
  \frac1{\beta_0\pi}\ln\frac1{1-\beta_0\as(Q)L},
\label{running as}
\end{equation}
where $\beta_0$ is given by
\begin{equation}
\beta_0=\frac{1}{4\pi} (\frac{11}{3} \;C_A - \frac{2}{3} N_f)~.
\end{equation}

The additional terms in the matrix $\Gamma$ are proportional to the
identity matrix.  There\-fore,  the diagonalization matrix $\R$ and
hence also $\wt{\H}$ (with a modified expressions for $\sigma^{(0)}$,
see above) and $\wt{\S}$ (\ref{tilde S}, \ref{tilde
H}) are unchanged. The only effect is to add a constant
$\frac{N_c^2-1}{4N_c}\rho(Y)$ onto both eigenvalues in the matrix $\D$
(\ref{D}). This results in an overall factor in the gap cross
section (see (\ref{fullall})). The full $LLQ_0$ gap cross section
(without non-global effects), $\sigma_{full}$, thus reads: 
\begin{align}
\sigma_{full} = \tilde{\sigma} \; \exp\left[ -\, \frac{N_c^2-1}{2N_c}
\rho(Y) \frac1{\beta_0\pi}\ln\frac1{1-\beta_0\as(Q)L}\right]
\label{sig full}
\end{align}
$\tilde{\sigma}$ is the same as $\sigma$ in the previous section
but with a running coupling and $\sigma^{(0)}$
replaced by $\sigma^{(0)}_{full}$. The non-global
logs \cite{DasSa1,DasSa2}
have not been resummed in the case of our observable, yet. An estimate of
the effect for a different gap definition can be found in \cite{AbSey03}.

It is worth noting that the structure of \eqref{sig full} is true for all flavours of
hard process.  However, it is only true if the gap definition is
azimuthally symmetric and one obtains a more complicated structure for
the gap definition used by H1 and ZEUS and calculated by Appleby and
Seymour \cite{AbSey03}, or for an $\eta$--$\phi$ `patch' as considered by Berger,
K\'ucs and Sterman \cite{BeKuSt}.

Fig.\ref{plo1_r_1} compares the DLLA (dotted line), the \pDLA (solid line) and the 
\pDLA with running coupling effects included (thick solid line). 
All results are taken in ratio with the full $LLQ_0A$ except that the lowest order
cross-section is always taken to be $\sigma^{(0)}$, even in $\sigma_{full}$ (i.e.
we are not concerned here with the small difference between $\sigma^{(0)}$ and 
$\sigma^{(0)}_{full}$). 
We use $R=1$ and  $\as(Q)=0.14$. The running of the 
coupling clearly plays an important role.

Fig.\ref{plo1_r_2} compares $\sigma$ with a running coupling to $\sigma_{full}$  
at different values of $L$ (again with the lowest order cross-section taken to
be $\sigma^{(0)}$ in both cases). The \pDL result increasingly differs from the full  
$LLQ_0$ result towards larger values of $L$. One obtains the full
cross-section to rather better than $10\%$ for $L < 7$.  

\EPSFIGURE[h]{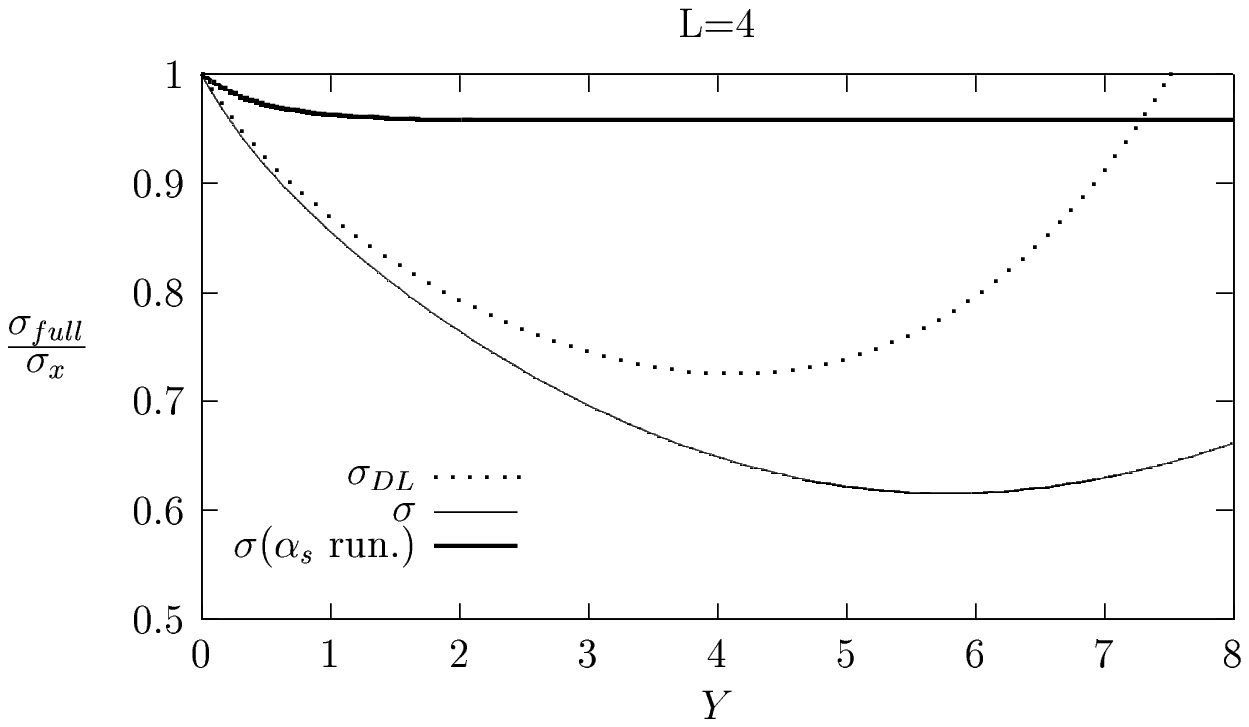,width=12cm}{The ratio of $\sigma_{full}/\sigma_x$
for $R=1$ \label{plo1_r_1}}

\EPSFIGURE[h]{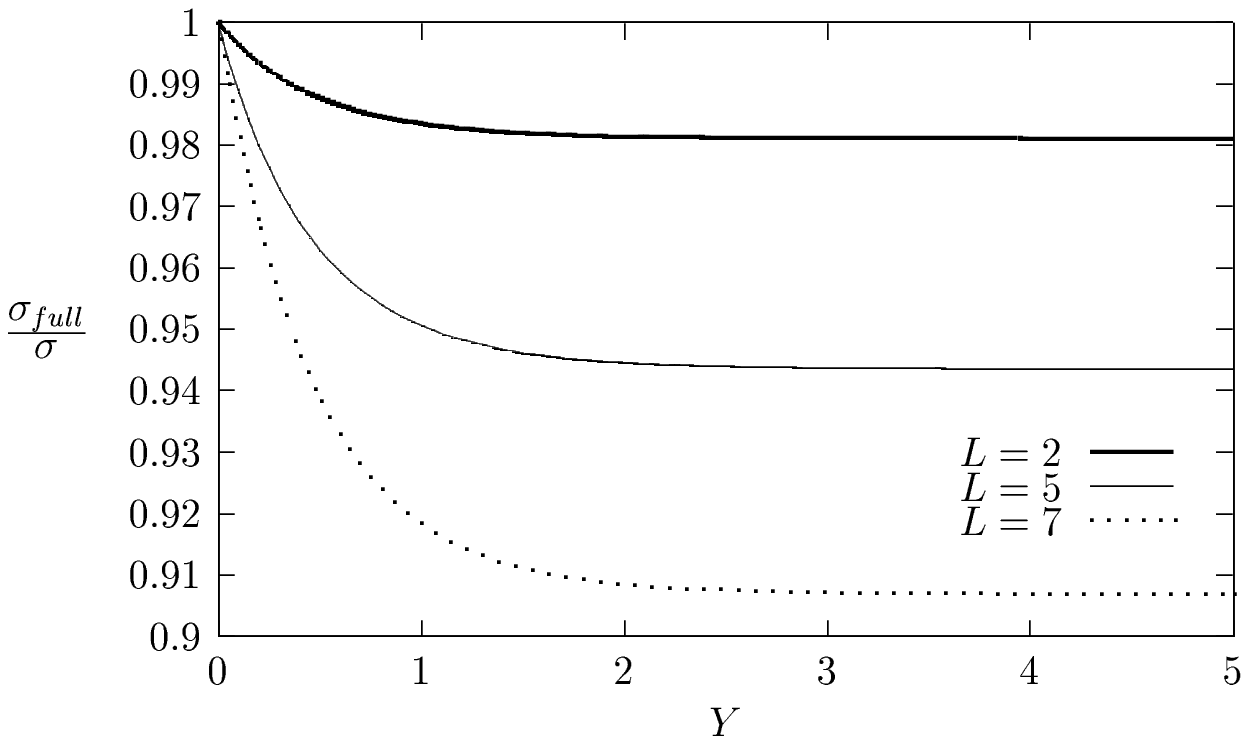,width=12cm}{The ratio of $\sigma_{full}/\sigma$
for $R=1$ and a running $\as$ \label{plo1_r_2}}


\subsection{Numerical results}
\EPSFIGURE[h]{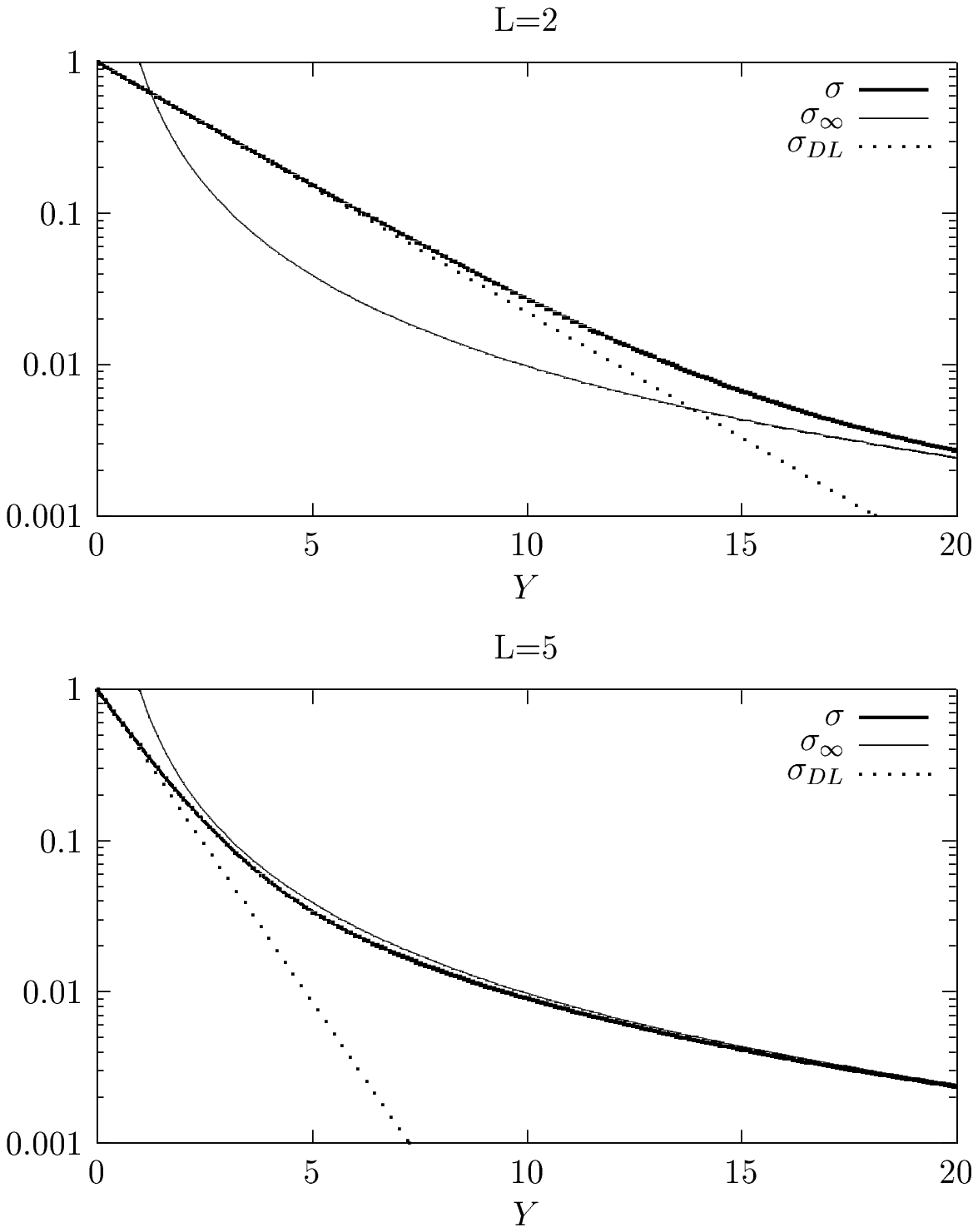}{The gap cross sections normalized to
$\sigma^{(0)}$ at different values of $L$ ($\as=0.2$). 
\label{numer}}

Before we continue to the BFKL cross section we numerically compare the cross
sections $\sigma$,  $\sigma_{DL}$ and $\sigma_\infty$ in (\ref{fullall}),
(\ref{no pi})  and (\ref{large Y}) using a
fixed coupling,  $\as=0.2$.  Fig.\ref{numer} shows these quantities
(divided by $\sigma^{(0)}$) for two values of $L$. $\sigma_\infty$, obtained by taking
$\sigma$ in the large-$Y$ limit,
is divergent at $Y\to 0$ and independent of $L$.
This is in contrast to $\sigma$ which falls with increasing $L$ at low and
intermediate values of $Y$. As a result, the very good agreement between
$\sigma$ and $\sigma_\infty$ at quite low values of $Y$, as illustrated in the
lower  pane of Fig.3, is somewhat accidental. The agreement becomes
worse again as $L$ increases. Of course, the two are always coincident
for large enough $Y$.
The cross section $\sigma_{DL}$ is strictly leading-$Y$  but, as demonstrated in
\eqref{limits}, it is not the large-$Y$ limit of
$\sigma$. Contrary to $\sigma_\infty$, the approximation of $\sigma$
by $\sigma_{DL}$ gets
better with decreasing $L$. So, including the imaginary non-leading-$Y$
parts of the loop integrals has completely changed the physical
result in the large $Y$- and/or $L$-limit.

\section{Matching with BFKL}
In this section we derive the elastic quark-quark scattering amplitude
via BFKL (colour singlet) 
exchange at any fixed order in $\as$.  Then, we determine the
terms it has in common with the \pDL amplitude calculated earlier.  We
discuss how to combine the two to provide a cross section that includes
the \pDL gap cross section and the first orders of the $2\to2$ BFKL
cross section without double counting.  Finally, we discuss a more
phenomenological way to combine them to all orders to provide a cross
section that agrees with each in its domain of validity and smoothly
interpolates the two.

We work in Feynman gauge and in $d=4-2\ep$ dimensions. We continue to
use the normalisation of $\A_8^{(0)}$ as given in \eqref{A8
norm} with a modified expression for $\sigma_0$ though. In $4-2\ep$
dimensions $\sigma_0$ \eqref{sig 0} is replaced by: 
\begin{align}
  \sigma^{(0)}_\ep = \mu^{2\ep} \left(\frac{4\pi\mu^2}{Q^2}\right)^\ep
  \frac{ \sigma^{(0)}}{\Gamma(1-\ep)}.
\end{align}

\subsection{The BFKL amplitude}
The elastic quark-quark scattering amplitude in
the leading-$Y$ approximation can be written as 
\begin{align}
\A_{BFKL}(Y, \q^2) = i s \as^2 \,(2 \pi \mu^2)^{2\ep}\, \frac{N_c^2-1}{N_c^2}  \int \frac{d^{2-2\ep}\ka_1 d^{2-2\ep}\ka_2}{\ka_1^2 (\ka_1-\q)^2}
 \;f(\ka_1, \ka_2, \q; Y),
\label{elas a}
\end{align}
where $f$ is given by the BFKL equation \cite{BFKL}
\begin{align}
\frac{\partial}{\partial Y}\, &f(\ka_1, \ka_2, \q; Y) =
\delta^{2-2\ep}(\ka_1-\ka_2) \delta(Y) +
\left[\omega(\ka_1^2)+\omega((\ka_1-\q)^2) \right] f(\ka_1, \ka_2, \q;
Y) \nonumber \\
-& \frac{\bar{\alpha}_s}{(2\pi)^{1-2\ep}} \int d^{2-2\ep}\el \left[
\frac{\q^2}{\el_1^2(\el-\q)^2} - \frac{1}{(\ka_1-\el)^2}
\left(\frac{\ka_1^2}{\el^2} +\frac{(\ka_1-\q)^2}{(\el-\q)^2} \right) \right] f(\el, \ka_2, \q; Y)
\label{BFKL}
\end{align}
with 
\begin{align}
\omega(\ka_1^2) = -\frac{1}{2} \frac{\bar{\alpha}_s}{(2\pi)^{1-2\ep}}
\int \frac{d^{2-2\ep}\el\: \ka_1^2}{\el^2(\el-\ka_1)^2}\:,\qquad
\bar{\alpha}_s=\frac{\as N_c}{\pi} \mu^{2\ep}.
\end{align}
In these equations we have $\q^2=Q^2$. We present two solutions for the
 elastic amplitude in the leading-$Y$ approximation. 
 First, we  solve  \eqref{BFKL} by iteration and derive the general
 expression for the amplitude at $\mathcal{O}(\as^n)$. The second
 solution has been derived in \cite{MoMaRy}; the result is written as
 a sum over conformal spins. 

In accordance with our previous notation, we denote by $f^{(n)}$ the
contribution to $f$ proportional to $\as^{n-1}$ (which therefore
contributes to the $n\!+\!1$-th order of the amplitude since the quark
impact factors each contain a factor $\as$).  The zeroth order
expression in \eqref{BFKL} is therefore $f^{(1)}$.
The momentum dependent part
of the elastic amplitude \eqref{elas a}   reads
\begin{align}
 \int &\frac{d^{2-2\ep}\ka_1 \d^{2-2\ep}\ka_2}{\ka_1^2 (\ka_1-\q)^2}
f^{(n)}(\ka_1, \ka_2, \q; Y) \nonumber \\
&= -\left(\frac{\bas Y}{n (2\pi)^{1-2\ep}}\right)
d^{2-2\ep}\ka_1 \;\d^{2-2\ep}\ka_2 \;\d^{2-2\ep}\el \;\Bigg\{ \nonumber\\
&{}\quad \frac{1}{2}
\frac{1}{(\el-\ka_1)^2}\left(\frac{1}{\el^2(\ka_1-\q)^2}+\frac{1}{\ka_1^2(\el-\q)^2
}\right)\;f^{(n-1)}(\ka_1, \ka_2, \q; Y)\nonumber \\
&\:+ \bigg[\frac{\q^2}{\el^2(\el-\q)^2\;\ka_1^2(\ka_1-\q)^2} \nonumber\\
&{}\qquad -\frac{1}{(\el-\ka_1)^2}\left(\frac{1}{\el^2(\ka_1-\q)^2} +
\frac{1}{\ka_1^2 (\el-\q)^2}\right) \bigg]\;f^{(n-1)}(\el,
\ka_2, \q; Y)\Bigg\} \label{step 1}\\
=\int& \frac{\d^{2-2\ep}\ka_1}{\ka_1^2(\ka_1-\q)^2} \:g^{(n)}(\ka_1, \q; Y)
\label{def g}
\end{align}
In the second term of \eqref{step 1} we substitute $\el
\leftrightarrow \ka_1$. The factor that multiplies  $f^{(n-1)}$ is
invariant under this transformation. The only effect therefore is that the arguments of
the functions $f^{(n-1)}$ in the first and in the second term become
equal and we can combine the two terms. Note that this is only
possible since the quark impact factors are constants. Performing the
$\ka_2$ integration we end up with the following
simple resursion formula for the function $g^{(n)}$ as defined in \eqref{def g} 
\begin{align}
&g^{(n)}(\ka_1, \q; Y) = \left(-\frac{\bas Y}{n (2\pi)^{1-2\ep}}\right)  \nonumber\\
&{}\quad \times \int \d^{2-2\ep}\el \bigg[\frac{\q^2}{\el^2(\el-\q)^2} - \frac{1}{2} \frac{1}{(\el-\ka_1)^2}\left(\frac{\ka_1^2}{\el^2} +
\frac{(\ka_1-\q)^2}{(\el-\q))^2}\right) \bigg]\;g^{(n-1)}(\ka_1,
 \q; Y) \nonumber \\
&=\left(-\frac{\bas Y}{2 n}\right)\nonumber \\
&{}\quad \times (4\pi)^\ep\; \frac{\Gamma(1+\ep)\Gamma(-\ep)^2}{\Gamma(-2\ep)} \left[ (\q^2)^{-\ep}
- \frac{1}{2} (\ka_1^2)^{-\ep} - \frac{1}{2} ((\ka_1-\q)^2)^{-\ep}\right] \;g^{(n-1)}(\ka_1,
 \q; Y) \label{new evo}\\
&g^{(1)} = 1.
\end{align}
To obtain the amplitude at $\mathcal{O}(\as^{n+1})$ we have to take the
bracket in (\ref{new evo}) to the power of
$n-1$ which gives rise to a double sum with two binominial coefficients and
we have to perform the $\ka_1$ integration. Our  result for the elastic
quark scattering amplitude is
\begin{align}
& \A^{(n)}_{BFKL} = i (-1)^{n} \A_8^{(0)} \as
\left(\frac{\as}{2\pi}\right)^{n-1}\;Y^{n-1}\;\frac{N_c^2-1}{8 (n-1)! \;N_c^{3-n}}
\;\left(\frac{\q^2}{4\pi\mu^2}\right)^{-n\ep} \nonumber \\
&\times \left(\frac{\Gamma(1+\ep)\Gamma(-\ep)^2}{\Gamma(-2\ep)}\right)^{n-1}\;
\sum_{k=0}^{n-1} \sum_{j=0}^k
\;\left(\begin{array}{c}n-1\\ k\end{array}\right)
\left(\begin{array}{c}k\\j\end{array}\right) (-1/2)^k\;
\frac{\Lambda(1+k)}{\Lambda(j)\,\Lambda(k-j)}\label{a all ord}
\end{align}
with
\begin{align*}
\Lambda(x) = \frac{\Gamma(1+x\ep)}{\Gamma(-(1+x)\ep)}\:,\qquad
\left(\begin{array}{c}n\\ k\end{array}\right) = \frac{n!}{k!(n-k)!}.
\end{align*}

To simplify further we perform the summation in
\eqref{a all ord}  for the first the few orders in
$\as$.  The result can be expressed in terms of functions $\gamma_n(\ep)$:
\begin{align}
\gamma_n(\ep) \equiv \frac{\Lambda(n+1)}{\Lambda(0)\Lambda(n)}
\end{align}
which obey also the following relation:
\begin{align}
&\int \frac{d^{2-2\ep}\ka\: \q^2}{(\ka-\q)^2 (\ka^2)^{1+n\ep}} =
\pi^{1-\ep}\, \gamma_n(\ep) \left(\q^2\right)^{-(1+n)\ep}\;,\quad n\in\mathrm{I\!N}_0.
\end{align}
Note that $\gamma_n(\ep)$ diverges as $1/\ep$.  
Suppressing the
argument of the functions $\gamma_n(\ep)$ we obtain:
\begin{align}
\A_{BFKL}^{(1)} &= -i \A_8^{(0)} \as \frac{N_c^2-1}{8N_c^2}
\left(\frac{\q^2}{4\pi \mu^2}\right)^{-\ep}\! \gamma_0 \quad\sim\frac{1}{\ep},\label{BFKL 1} \\
\A_{BFKL}^{(2)} &= i \A_8^{(0)} \frac{\as^2}{(2\pi)} \frac{N_c^2-1}{8N_c}\;Y\;
\left(\frac{\q^2}{4\pi \mu^2}\right)^{-2\ep} \!
\gamma_0\left(\gamma_0-\gamma_1\right) \quad\sim\frac{1}{\ep^2},\label{BFKL 2} \\
\A_{BFKL}^{(3)} &= -i \A_8^{(0)} \frac{\as^3}{(2\pi)^2}
\frac{N_c^2-1}{16}\;
Y^2\;\left(\frac{\q^2}{4\pi\mu^2}\right)^{-3\ep} \nonumber \\
&{}\qquad
\gamma_0\left(\gamma_0(\gamma_0-\gamma_1) - \gamma_1(\gamma_0-\gamma_2)
+ \frac{1}{2} \gamma_2(\gamma_0-\gamma_1 )\right) \quad\sim\frac{1}{\ep^3} \label{BFKL 3}
\end{align}
Here, we have also noted the leading behaviour of the amplitudes as $\ep\to 0$.

The all orders  leading-$Y$ result for the elastic quark-quark amplitude,  $\A_{BFKL}$, is
written as a power series in $\as$ in our approach.  
In contrast to this it can also be written as a sum over conformal
spins \cite{MoMaRy}
\begin{align}
 \A_{BFKL}&= -i \A_8^{(0)}\,\frac{\as}{2\pi}\frac{N_c^2-1}{ N_c^2} \nonumber \\
&\times\sum_{m=-\infty}^{\infty}
\;\int d\nu \left\{ \frac{\nu^2+m^2}{[\nu^2+(m-1/2)^2][\nu^2+(m+1/2)^2]}
\exp\left[\omega_{2m}(\nu)\,Y\right]\right\}
\end{align}
with
\begin{align}
\omega_n(\nu) = \frac{N_c\as}{\pi}\left[2
\psi(1)-\psi(1/2+|n|/2+ i\nu)-\psi(1/2+|n|/2-i\nu)\right]
\end{align}
and $\psi(z)=1/z \;\partial\Gamma(z)/\partial z$.  It is worth
emphasizing that this all-orders expression is finite, even though its
expansion to any finite order is divergent, as seen above.  Physically,
this is because the divergent terms exponentiate to zero.  This property
will be crucial in constructing an all-orders matched cross section
below.

In the limit $Y\to
0$ the BFKL $2\to 2$ cross section  reduces to \cite{MoMaRy} 
\begin{align}
\left.\sigma_{BFKL}\right|_{Y\to 0} = N_c^2 \left|\A_{BFKL}\right|^2_{Y\to 0} = \sigma^{(0)}
\frac{N_c^2-1}{N_c^4} \frac{\pi^2}{Y^2}.
\label{sig BFKL lim}
\end{align}
This behaviour will also be important in the all-orders combination of
BFKL with the gap cross section.

\subsection{Singlet exchange in the \pDLA}
As the first step towards a matching of our gap cross section to BFKL
we extract from  the elastic amplitude in the \pDLA its singlet
exchange part taken in the leading-$Y$ approximation. To be more
precise, we extract from the resummed
 elastic amplitude \eqref{result ampl} 
the singlet exchange contribution $\A_1^{(n)}(Q_0)$ at fixed order in
$\as$. Keeping in this expression only the terms leading
in $Y$ we obtain an expression we term $\A_{1,S}^{(n)}(Q_0)$. The subscript stands for
`strongly ordered' since at
$Q_0=0$ this amplitude differs from the leading-$Y$ (BFKL)
amplitude in that it is calculated in the strongly ordered
approximation. 
For the first few values of $n$ we have:
\begin{align*}
&(\vec\A^{(0)}(Q_0))_{1,S} = 0,\quad (\vec\A^{(1)})_{1,S} =
i\pi\,\frac{N_c^2-1}{4N_c^2}\,\A_8^{(0)} \,(-\frac{\as}{\pi} L)\equiv a\; (-\frac{\as}{\pi} L)\\
&(\vec\A^{(2)}(Q_0))_{1,S} = \frac{1}{2!} (-\frac{\as}{\pi} L)^2 (\Gamma^2
\vec\A^{(0)})_1 =  \frac{1}{2} a\, (-\frac{\as}{\pi} L)^2
\,(\frac{N_c}{2}Y),\\
&(\vec\A^{(3)}(Q_0))_{1,S} = \frac{1}{3!}  (-\frac{\as}{\pi} L)^3 (\Gamma^3
\vec\A^{(0)})_1 =  \frac{1}{3!} a (-\frac{\as}{\pi} L)^3 \,(\frac{N_c}{2}Y)^2\\  
\end{align*}
The general result for $n\ge 1$ can be written as
\begin{align}
\label{singl soa}
\A^{(n)}_{1,S}(Q_0) = i\, \A^{(0)}_8\;\frac{N_c^2-1}{2N_c^3}\;\frac{\pi}{Y} \;
\frac{1}{n!} \left(-\frac{N_c \as}{2\pi} \;Y L\right)^n 
\end{align}
and one obtains for the resummed ($LY$) singlet exchange amplitude
\begin{align}
\A_{1,S}(Q_0)= -i\,\frac{N_c^2-1}{2N_c^3}\;\frac{\pi}{Y} \; \A^{(0)}_8
\left[ 1 - \exp\left(-\frac{N_c \as}{2 \pi} \;Y L\right) \right].
\label{singl resum}
\end{align}
In the limit of large $Y$ or $L$ the exponential vanishes and we 
obtain a finite result for the all-orders amplitude $\A_{1,S}(0)$.
\begin{align}
\A_{1,S}(0) = \left.\A_{1,S}(Q_0)\right|_{L\to \infty} = -i\,\frac{N_c^2-1}{2N_c^3}\;\frac{\pi}{Y} \; \A^{(0)}_8.
\end{align}
Using \eqref{A8 norm}  the corresponding cross section
$\sigma_{S}$ reads
\begin{align}
\sigma_{S} = \C_1^2\;|\A_{1,S}(0)|^2 = 
\;\frac{\sigma^{(0)}}{\C_8^2 |\A_8|^2}\C_1^2\;|\A_{1,S}(0)|^2 =
\sigma^{(0)}\;\frac{N_c^2-1}{N_c^4}\; \frac{\pi^2}{Y^2}.
\end{align}
This is exactly the large $Y$ limit of our gap cross
section \eqref{large Y} and the small $Y$ limit of the BFKL cross
section \eqref{sig BFKL lim}
\begin{align}
\left.\sigma_{BFKL}\right|_{Y\to 0} = \sigma_{S} = \left.\sigma\right|_{Y\to \infty}
\end{align}
We will exploit this remarkable fact to construct an all orders
combined cross section further below. 

\subsection{Fixed order matching to BFKL}
In order to combine the $2\to 2$ BFKL cross section with the gap
cross section $\sigma$ we have to ensure that a) at each order in $\as$
the $\ep$-poles in the BFKL cross section are cancelled and b) we have
avoided any double counting. The only contribution to $\sigma$ which
also appears in the BFKL cross section  is $\A^{(n)}_{1,S}(0)$ \eqref{singl soa}.
Let us keep $Q_0\ne 0$ for the moment and consider the generalization
of the amplitude \eqref{singl soa} to $d=4-2\ep$ dimensions:  
\begin{align}
\A^{(n)}_{1,S}(Q_0) =&i \A^{(0)}_8
\frac{N_c^2-1}{2N_c^3}\;\frac{\pi}{Y} \,\frac{1}{n!} \left( -\frac{N_c\as}{\pi} Y\,
\frac{(4\pi\mu^2)^\ep}{\Gamma(1-\ep)} \int_{Q_0}^Q
\frac{d k }{k^{1+2\ep}} \right)^n
\label{ImA d}
\end{align}
We have calculated this amplitude explicitly in $d$ dimensions only up
to $\mathcal{O}(\as^3)$ accuracy (see appendix), but using the
well-known fact that the eikonal Feynman rules are unchanged in $d$
dimensions, so only the phase space integrals become $d$-dimensional, it
is straightforward to see that (\ref{ImA d}) will hold to any order.
Now we can set $Q_0=0$ and obtain the
elastic scattering amplitude in the leading-$Y$
approximation and in the
approximation of strongly ordered transverse momenta (indicated by the
subscript '$S$'):
\begin{align}
\A^{(n)}_{1,S}(0) = &\, i \A^{(0)}_8 \as
\frac{N_c^2-1}{2N_c^2}\;\frac{1}{n!} \left(\frac{N_c \as}{\pi} Y\right)^{n-1}
\left(\frac{1}{2\ep} \;\frac{1}{\Gamma(1-\ep)} \right)^n \left(\frac{\q^2}{4 \pi
\mu^2}\right)^{-n \ep}
\label{A1S}
\end{align}
Now, we calculate the difference of the BFKL amplitude \eqref{a all
ord} and its strongly ordered approximation \eqref{A1S} for the first orders in $\as$:
\begin{align}
\delta^{(n)}\equiv\A_{BFKL}^{(n)}-\A^{(n)}_{1,S}(0)
\label{delta}
\end{align}
\begin{align}
& \delta^{(1)}= - i \A_8^{(0)} \as
\frac{N_c^2-1}{2N_c^2} \zeta(3)\,\ep^2 + \mathcal{O}(\ep^3)
\label{cancel 1}\\
& \delta^{(2)} = i \A_8^{(0)} \as^2
\frac{7}{8} \,\frac{Y}{\pi}\,\frac{N_c^2-1}{N_c}  \zeta(3) \ep +
\mathcal{O}(\ep^2) \\
& \delta^{(3)} = - i \A_8^{(0)} \as^3
\frac{7}{24} \,\frac{Y^2}{\pi^2}\,(N_c^2-1)  \zeta(3)  +
\mathcal{O}(\ep)
\label{cancel 2}\\
& \delta^{(4)} = \mathcal{O}(1/\ep)
\end{align}
Here, we have set $\ep=0$ in $\A_8^{(0)}$. $\zeta(x)$ is  the
Riemann $\zeta$-function. The leading pole in the BFKL
amplitude is captured by the strong ordering approximation. And, not
surprisingly, the strongly ordering approximation cannot provide all
$\ep$-poles; at $\mathcal{O}(\as^5)$ the difference of the two
expressions starts to become divergent. Working in
the \pDLA (or in the $LLQ_0A$)  will allow us therefore only to
include the BFKL cross section to a fixed order of $\as$. Note that  the
smallest power of $\ep$ in $\A^{(1)}_{BFKL}$ is $\ep^{-1}$, in
$\A^{(2)}_{BFKL}$ it is $\ep^{-2}$ etc. 
(see \eqref{BFKL 1}-\eqref{BFKL 3}). Remarkably, the strong ordering
approximation therefore  agrees with the full amplitude not only at  the
smallest power of $\ep$ but rather at the three smallest powers.  

Now we come  to our original goal to include the BFKL cross
section in our gap cross section.
We start from the Theorem stated above and work at fixed order in $\as$. Denoting the production
amplitude (in the \pDLA, including color
factor) for more than 2 particles by $\B(Q_0)$ the theorem reads 
\newcommand{\SaQ}{\A_{1,S}(Q_0)}
\newcommand{\Sa}{\A_{1,S}(0)}
\newcommand{\FO}{\A_{8}(0)}
\newcommand{\FOQ}{\A_{8}(Q_0)}
\newcommand{\FSQ}{\A_{1}(Q_0)}
\newcommand{\FS}{\A_{1}(0)}
\newcommand{\mbQ}{\B(Q_0)}
\newcommand{\FL}{\A_{BFKL}}
\begin{align}
\sigma^{(k)} = |\FS|^2\C_1^2 + |\FO|^2\C_8^2 + |\mbQ|^2 =
|\FSQ|^2\C_1^2 + |\FOQ|^2\C_8^2  
\label{theorem}
\end{align}
where the squares are to be read symbolically representing the sums over
$\A^{(n)*}\A^{(m)}$ (and $\B(Q_0)$ , respectively). 
In this paper, we wish to also include the colour singlet BFKL $2\to 2$
amplitude $\A^{(k)}_{BFKL}$.\footnote{We do not aim here to include those
BFKL logarithms generated as a result of real emissions.}
However, 
$\A_1(0)$ also includes terms subleading in $Y$ , whereas the BFKL
amplitude is leading-$Y$. We therefore have to keep these subleading
terms; they are given by $(\A_1(0)-\A_{1,S}(0))$ since 
$\A_{1,S}(0)$ is calculated in the leading-$Y$ approximation.
 Omitting indices for the moment we therefore define the following
fixed order gap cross section
\begin{align*}
\sigma^{(k)}_{gap} &\equiv |\FL + \FS - \Sa |^2 \C_1^2+
|\FO|^2\C_8^2 + |\mbQ|^2\\ 
&=|\FS+\delta|^2\C_1^2 + |\FO|^2\C_8^2 + |\mbQ|^2\\
&=|\FS|^2\C_1^2 + |\FO|^2\C_8^2 + |\mbQ|^2 + (2
\Re\left[\FS\delta^*\right] + |\delta|^2)\C_1^2\\
&=\sigma^{(k)} + (2 \Re\left[\FS \delta^*\right] + |\delta|^2)\C_1^2
\end{align*}
where, in the last  line we have invoked the theorem \eqref{theorem}. 
Exploiting the fact that $\delta$ \eqref{delta} is purely imaginary and
showing the indices again we end up with
\begin{align}
\sigma^{(k)}_{gap}=\sigma^{(k)} + \Delta^{(k)}\:,\quad
\Delta^{(k)}&= N_c^2 \sum_{m+n=k} \left[ 2
\Im\A^{(m)}_1(0)\cdot(-i \delta^{(n)}) + \delta^{(m)}\delta^{(n)*}\right]\label{albrecht}\\
\delta^{(n)}&=\FL^{(n)}-\A^{(n)}_{1,S}(0)\nonumber
\end{align}
It now is easy to see at which order of $\as$ the function $\Delta^{(k)}$ and hence
$\sigma^{(k)}_{gap}$ starts to be
divergent. $\delta^{(n)}$ is finite for $n\le 3$, see \eqref{cancel
1}-\eqref{cancel 2}. $\Im \A_1^{(1)}(0)$ only has a $1/\ep$-pole; this gets
multiplied by a constant, namely $\delta^{(3)}$ \eqref{cancel 2},  for
the first time in $\Delta^{(4)}$.  Hence, $\Delta^{(k)}$ is  
expected  to include $\ep$-poles for $k\ge 4$.
To calculate $\Delta^{(2)},\Delta^{(3)}$ we need
$\A^{(1)}_1(0), \A^{(2)}_1(0)$ which are given in the appendix
($\A_1^{(3)}$  is not needed since $\A_1^{(0)}$ is zero). Comparing with
\eqref{A1S} it turns out that these amplitudes have no subleading (in $Y$) parts,
\begin{align*}
\Im\A^{(k)}_1(0) = \Im\A^{(k)}_{1,S}(0)\:,\quad k=1,2.
\end{align*}
We can therefore use the leading-$Y$ expression $\A^{(m)}_{1,S}(0)$ instead of $\A^{(m)}_1(0)$
in \eqref{albrecht} and inserting the definition of $\delta^{(n)}$ we arrive
at:
\begin{align}
\Delta^{(k)} &= N_c^2 \sum_{m+n=k} \left[ 2
\Im\A^{(m)}_{1,S}(0)\cdot(-i \delta^{(n)}) +
\delta^{(m)}\delta^{(n)*}\right] ,\quad  k\le 3 \nonumber\\
&=N_c^2 \sum_{n+m=k} \left[ \Im\A_{BFKL}^{(n)} 
\Im\A_{BFKL}^{(m)} - \Im\A_{1,S}^{(n)}(0)\, \Im\A_{1,S}^{(m)}(0)
\right]
\label{Delta p}
\end{align} 
 Using (\ref{BFKL 1}),(\ref{BFKL 2}) and
(\ref{ImA d}) in (\ref{Delta p}) we obtain:
\begin{align}
\Delta^{(2)} &= \mathcal{O}(\ep)\\
\Delta^{(3)} &= \sigma^{(0)} \frac{3}{2} \as^3 \frac{N_c^2-1}{N_c}
\frac{Y}{\pi} \zeta(3) \label{Delta3}\\
\Delta^{(4)} &= \mathcal{O}(1/\ep) 
\end{align}
Since $\Delta^{(k)}$ diverges for $k\ge 4$ our final gap cross section includes the \pDL 
 result and the BFKL cross section up to
$\mathcal{O}(\as^5)$,
\begin{align}
\sigma_{gap} &\equiv\Delta^{(2)} + \Delta^{(3)} + \sum_{k=0}^{\infty} 
\sigma^{(k)}\nonumber\\
&= \sigma +  \Delta^{(3)}\label{sig comb}
\end{align}
where the resummed cross section $\sigma$ is given by
\eqref{fullall}.

 $\Delta^{(2)}$ corresponds to the squared 1-loop
amplitude; the fact that $\Delta^{(2)}$  is zero means that the
gap cross section  $\sigma$ fully includes the leading order
BFKL cross section ($\sigma$  also includes the terms that cancel the
corresponding divergence). 

Note that we could include in  $\sigma$ terms that are suppressed in the high
energy limit (including  non-global logarithms). This
would not affect the value of $\Delta^{(3)}$ needed for a combined 
$LLQ_0 + $~BFKL cross section. Therefore, we
can use $\sigma_{full}$, (\ref{sig full}), instead of $\sigma$
in \eqref{sig comb} and obtain the cross section that combines the
conventional (partonic) gap cross section \cite{OdSt98, Od00} with the
first orders of the BFKL result (not taking into account the
non-global logarithms).

Fig.\ref{numer BFKL} compares the combined gap cross section
$\sigma_{gap}$ (\ref{sig comb}) with $\sigma$. We use a fixed 
coupling $\as=0.2$. Note that part of the BFKL cross
section is included in $\sigma$. Fig.\ref{numer BFKL} therefore shows
the difference between the full (up to $\mathcal{O}(\as^5)$) and the
partial inclusion of BFKL.  The importance of
BFKL grows towards larger values of $Y$ and/or $L$ since $\sigma$ decreases whereas
the BFKL cross section is independent of $L$ and grows with $Y$.  

\EPSFIGURE[h]{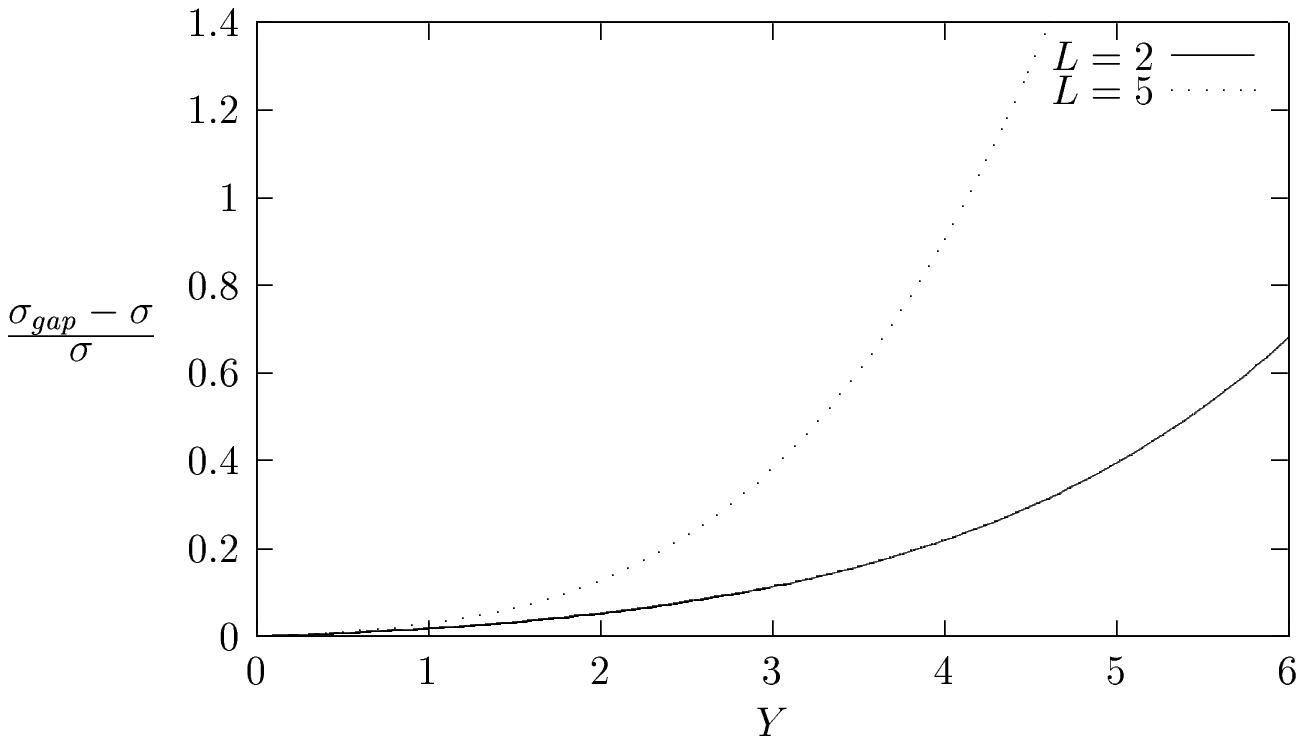}{The fixed-order combined gap cross section compared to
$\sigma$ ($\as=0.2$) (normalised to $\sigma^{(0)}$).\label{numer BFKL}}

The inclusion of the first orders of the BFKL cross section in the gap cross section is
found to change the result by a factor $\sim 2$ for $Y=4$ and $L=5$. 
At the
Tevatron in Run II jets with $p_T\sim 30$ GeV and rapidity separation
of $\Delta \eta=6$ (i.e. $Y=4$) can be expected. For  $Q_0=2$ GeV such jets
corresponds to $L=5.4$. The impact of the singlet exchange cross
section might therefore be observable at the Tevatron. At the LHC
values of $Y=6$ and $L>5$ are most probably accessible, enhancing the
effect of BFKL even more. 

Fig.\ref{gap BFKL} shows our
results together with the resummed BFKL cross section \cite{MoMaRy}. The BFKL
cross section should be appropriate for large enough $Y$ whilst $\sigma_{full}$ 
is the most reliable for small values of $Y$. A combined cross
section should therefore smoothly interpolate between the large-$Y$
BFKL cross section and  $\sigma$ at small $Y$. 

\EPSFIGURE[h]{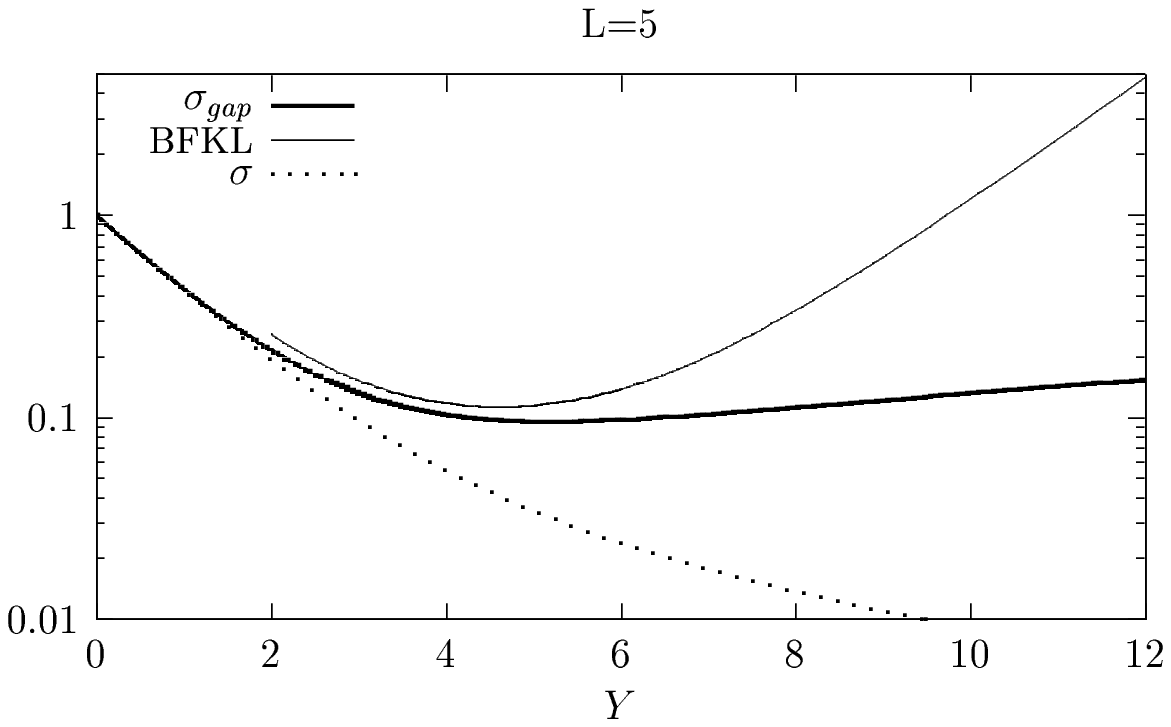}{The fixed-order combined gap cross section compared to the
full BFKL result  ($\as=0.2$)\label{gap BFKL}}

\subsection{All orders matching to BFKL}

In the previous section we showed that an order-by-order matching of the
$LLQ_0$ and BFKL results can only work for the first few orders, because
beyond the fourth order, the sub-leading poles in the amplitude do not
match and the combination is divergent.  We now show that, using the
fact that the all-orders results for $\A_{BFKL}$ and $\A_{1,S}(0)$ are
finite, and that they are identical at small $Y$, it is possible to
construct an all-orders cross section that does smoothly interpolate the
$LLQ_0$ and BFKL results, agreeing with each in its region of validity.
In fact we can construct several different matched cross sections that
all fulfil these requirements while differing slightly in their
predictions in the intermediate region.  We show numerical results from
all three as a measure of the uncertainty inherent in the matching
procedure.  We emphasize that these are all phenomenological approaches
that cannot be proved order by order, since they are based on all-orders
summations of divergent expressions.

\subsubsection{Simple Matching}

In the previous section, we showed that the combined gap cross section
could be written as
\begin{equation}
  \sigma_{gap} = \sigma +
  N_c^2 \left[ 2\Im\A_1(0)\cdot(-i \delta) + |\delta|^2\right],
  \label{simple matching}
\end{equation}
where we have dropped the superscripts, implying that we are summing
each expression to all orders.  Although all the expressions appearing
in the square brackets in (\ref{simple matching}) are divergent at
every order of perturbation theory, they all have finite all-orders
expressions.  In particular, $\A_1(0)$, the $LLQ_0$ singlet amplitude
for fixed $Y$ and $L\to\infty$, is zero.  We therefore have
\begin{equation}
  \sigma_{gap} = \sigma + N_c^2 |\delta|^2.
  \label{simple matching2}
\end{equation}
This expression achieves our goal of having a smooth matching of the two
all-orders cross sections, in that for small and large $Y$ it agrees
with the $LLQ_0$ and BFKL cross sections respectively.  For small $Y$
(and any $L$, since they are both $L$-independent), we already showed
that $\A_{BFKL}$ and $\A_{1,S}(0)$ become identical.  Therefore $\delta$
becomes zero and $\sigma_{gap}$ becomes equal to $\sigma$.  For large
$Y$ and any $L$, $\sigma$ and $\A_{1,S}(0)$ both become zero and hence
$\sigma_{gap}$ becomes equal to the BFKL cross section.

\subsubsection{Cross Section Matching}
In the previous section, we also showed that for the orders for which it
is finite, (\ref{simple matching}) is identical to
\begin{eqnarray}
  \label{sigma matching}
  \sigma_{gap} &=& \sigma +
  N_c^2 \left[ \Im\A_{BFKL}\Im\A_{BFKL} - \Im\A_{1,S}(0)\, \Im\A_{1,S}(0)
    \right] \\
  &\equiv& \sigma + \sigma_{BFKL} - \sigma_{S}.
\end{eqnarray}
We take this as the definition of another matching scheme, the cross
section matching scheme.  It has a very simple interpretation: the
matched cross section is equal to the sum of the $LLQ_0$ and BFKL cross
sections, with the double-counted terms subtracted off.  Note that
according to the discussion in the previous section, this can be written
in an identical form to (\ref{simple matching}), but with $\A_1(0)$
replaced by $\A_{1,S}(0)$.  Since $\A_1$ becomes equal to $\A_{1,S}$ for
sufficiently large $Y$, and the $\delta$ term that these terms multiply
becomes zero for small $Y$, these expressions are equally good within
their regions of applicability.

This expression therefore also achieves our goal of having a smooth
matching of the two all-orders cross sections, in that for small and
large $Y$ it agrees with the $LLQ_0$ and BFKL cross sections
respectively.

\subsubsection{Amplitude Matching}
Finally, inspired by the form of the cross section matching scheme, we
consider a similar matching, but at the amplitude level,
\begin{eqnarray}
  \label{A matching}
  \sigma_{gap} &=& \frac14(N_c^2-1)|\A_8(Q_0)|^2 + N_c^2|\A_{1,gap}(Q_0)|^2, \\
  \A_{1,gap}(Q_0) &\equiv& \A_{1}(Q_0)+\delta \\
  &=& \A_{1}(Q_0)+\A_{BFKL}-\A_{1,S}(0).
\end{eqnarray}
That is, we write the singlet amplitude as the sum of the $LLQ_0$ and
BFKL singlet amplitudes, with the double-counted terms subtracted off.
Expanding this expression and rewriting it in terms of $\delta$, we
obtain
\begin{equation}
  \sigma_{gap} = \sigma +
  N_c^2 \left[ 2\Im\A_1(Q_0)\cdot(-i \delta) + |\delta|^2\right].
  \label{A matching2}
\end{equation}
That is, an identical expression to (\ref{simple matching}) but with
$\A_1(0)$ replaced by $\A_1(Q_0)$.  Since $\A_1(Q_0)$ becomes
$Q_0$-independent for sufficiently large $Y$, and the $\delta$ term that
these terms multiply becomes zero for small $Y$, these expressions are
equally good within their regions of applicability.

This expression therefore also achieves our goal of having a smooth
matching of the two all-orders cross sections, in that for small and
large $Y$ it agrees with the $LLQ_0$ and BFKL cross sections
respectively.

\subsection{Numerical Results}
We show numerical results of all three schemes in Figs.~\ref{gap simple}, 
\ref{gap cross sec}, \ref{gap amplitude} for
$L=2$, 5 and 6.  We see that indeed they all achieve the goal of
matching the two cross sections in the small and large $Y$ limits and
the differences are not large in between.  The very close agreement that one sees at
$L=5$ is somewhat coincidental.

\EPSFIGURE[p]{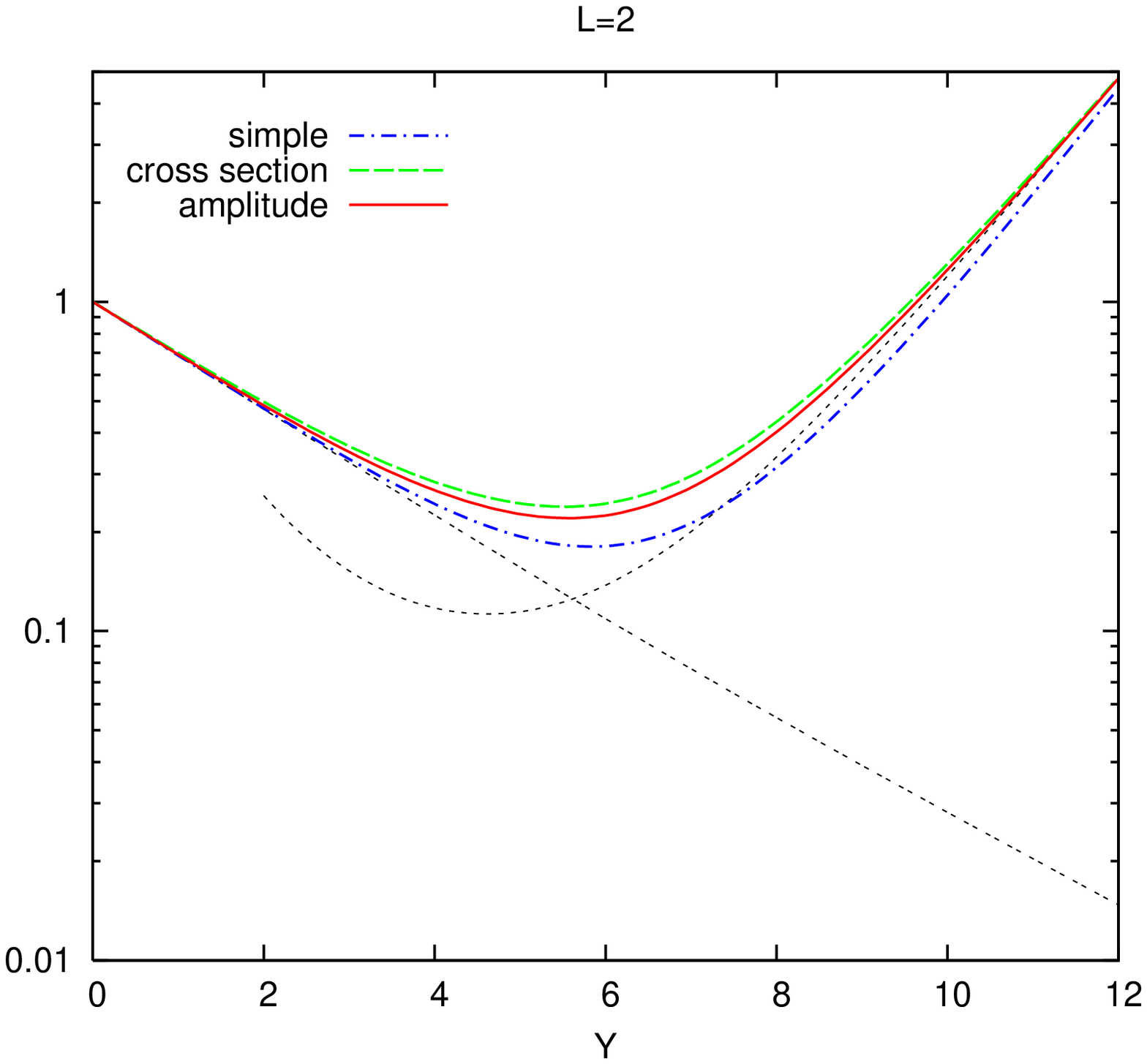,width=14.6cm}{The gap cross section in the
three matching schemes for $L=2$ ($\as=0.2$) compared to
$\sigma_{BFKL}$ (dots) and $\sigma$ (double-dots)\label{gap simple}}

\EPSFIGURE[p]{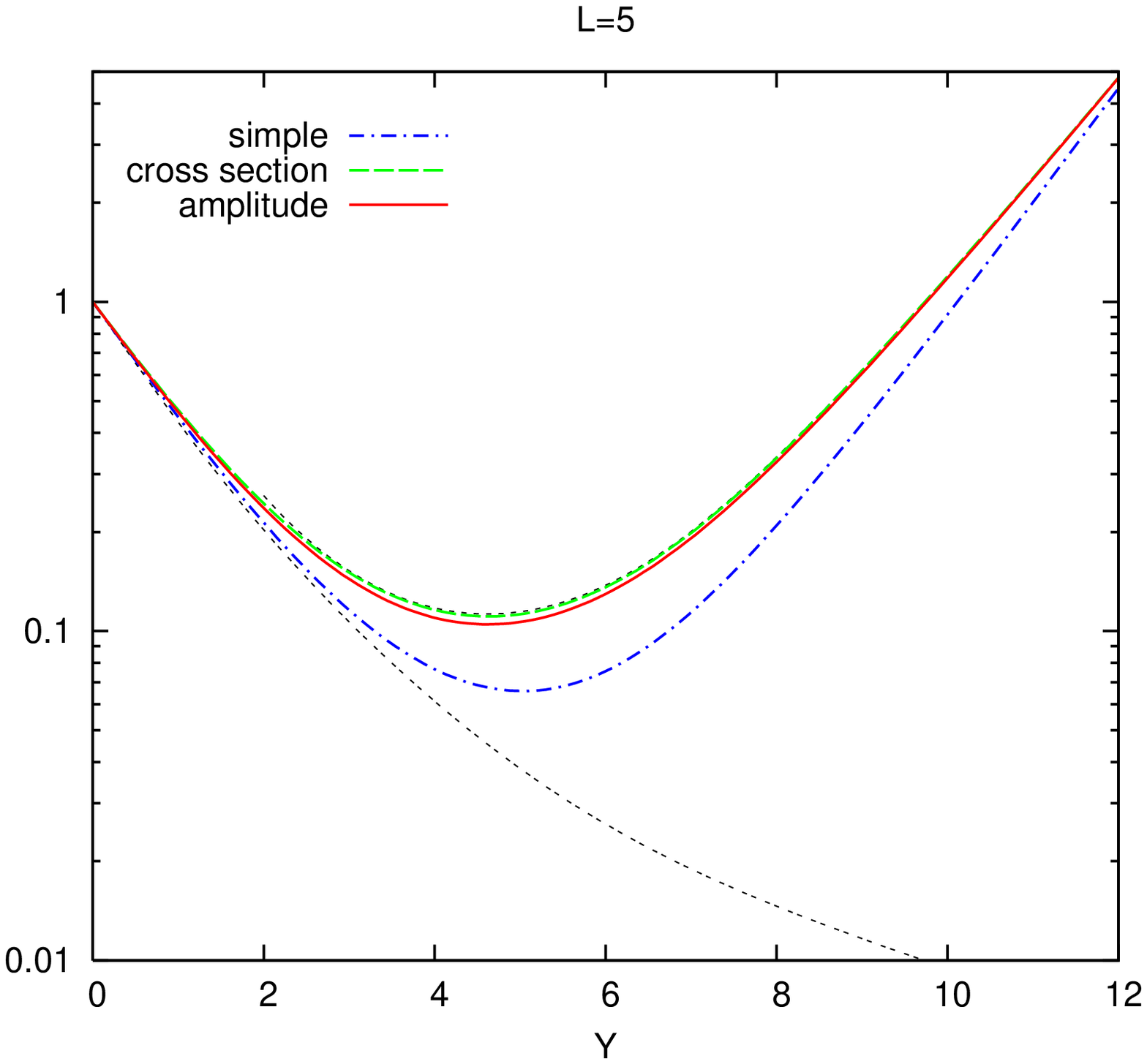,width=14.6cm}{The gap cross section in the
three matching schemes for $L=5$ ($\as=0.2$) compared to
$\sigma_{BFKL}$ (dots) and $\sigma$ (double-dots)\label{gap cross sec}}

\EPSFIGURE[p]{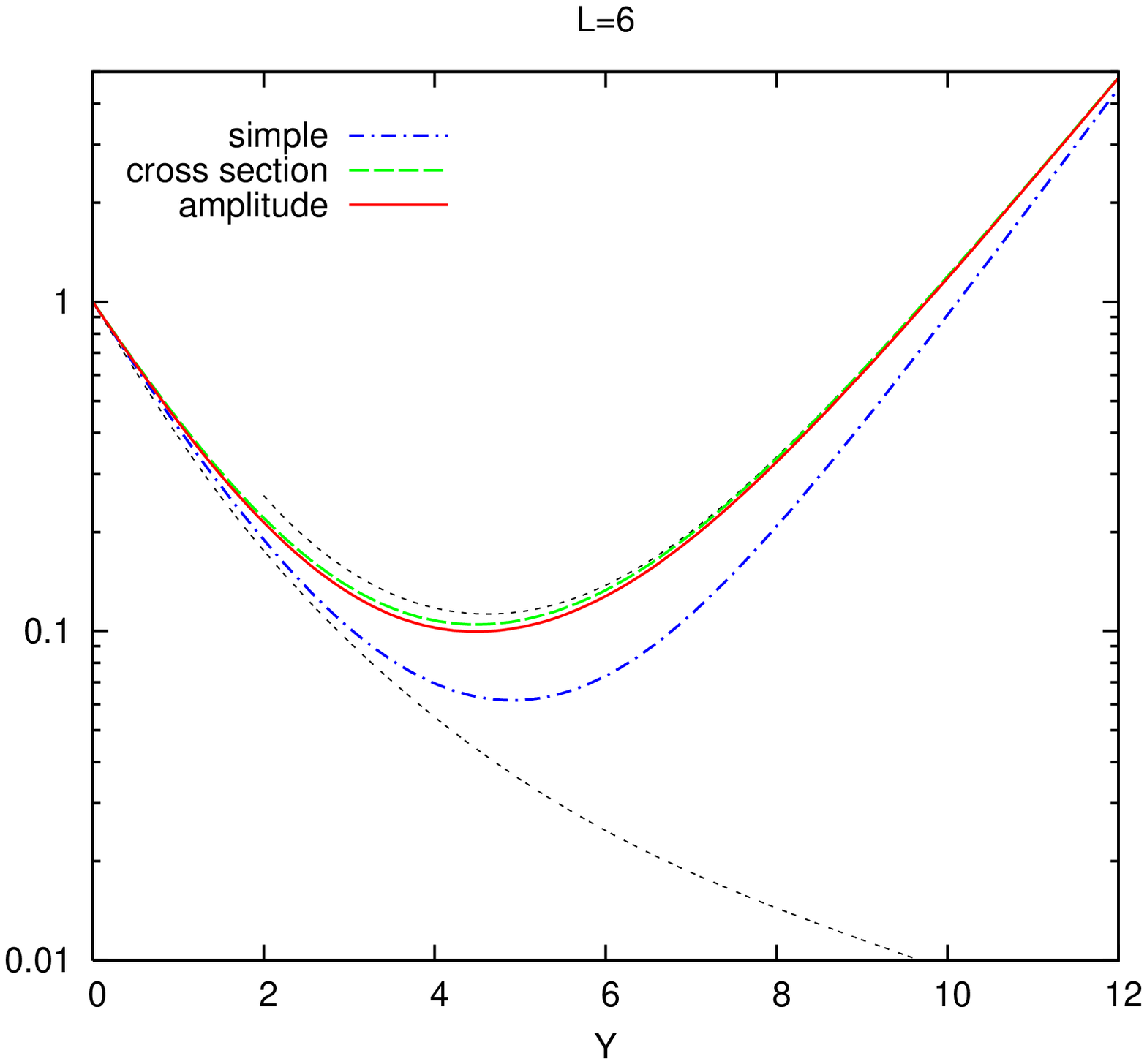,width=14.6cm}{The gap cross section in the
three matching schemes for $L=6$ ($\as=0.2$) compared to
$\sigma_{BFKL}$ (dots) and $\sigma$ (double-dots)\label{gap amplitude}}

We stress again that all three approaches are formally equally valid and
equally phenomenological, in the sense that they cannot be justified by
a finite order-by-order expansion.  Only with a better understanding of
the real-emission contributions in the high energy limit, which also
have finite all-orders expressions but cancel these divergences order by
order, can a more formally-justified matching be set up.  We reserve
this for future work.

\section{Conclusion}
Working in the high energy limit we have calculated the (partonic)
cross section for the production of two jets distant in rapidity 
and with limited transverse energy flow into the
region between the jets. Besides the DLL terms, we have summed
terms sub-leading in $Y$ stemming from the imaginary parts of the loop
integrals. This allowed us to identify those contributions that
are also included in the fixed order $2 \to 2$ BFKL cross
section. We have shown that the leading order BFKL cross section,
corresponding to the squared 1-loop amplitude, is entirely contained
in the $\pi^2 DLLA$. We derived an expression that consistently includes
the terms of the $LLQ_0$ series and the BFKL series to
$\mathcal{O}(\as^5)$ accuracy without double counting. In the 
$LLQ_0A$, the inclusion of higher orders of the BFKL cross section
in this way is not possible since it implies a divergent cross
section. 

We have also studied several phenomenological ``all order'' matching schemes that
effectively interpolate between the $LLQ_0$ and BFKL results.  Although
they all yield similar results, the differences between them cannot be
resolved without further work, specifically understanding the role of
real-emission contributions in the high energy limit.

In this paper we have made a first step towards the unification of the two
main approaches to the ``jet--gap--jet'' process.  The question remains as to
how they can be fully combined.

\appendix

\section{Elastic Scattering Amplitudes in \boldmath$d$ Dimensions}
We calculate the elastic quark-quark amplitude in the high energy
limit  in the  approximation of  strongly-ordered
transverse momenta of the $t$-channel gluons. We work in the eikonal
effective theory.  The transverse momentum of the
softest gluon is taken to be larger than $Q_0$. Since we need the singlet exchange
amplitude for $Q_0=0$ we consider $d=4-2\ep$ dimensions. We work in
Feynman gauge. Denoting the color indices of the initial and final
state quark pair by ${ik}$ and ${jl}$, respectively, the colour
operators for singlet and octet exchange are
\begin{align*}
\C_1 = \delta_{ij} \delta_{kl}\quad ,\qquad\C_8 = \frac{1}{2} ( \delta_{il} \delta_{jk} - \frac{1}{N_c}
\delta_{ij} \delta_{kl}).
\end{align*}
The tree level amplitude reads:
\begin{align}
\A^{(0)}\C_8 = -g_s^2\mu^{2\ep}\frac{2\shat}{Q^2}\C_8
\end{align}
\subsection{The one loop amplitude}
There are two Feynman diagrams, the box and the crossed box. In each
of them there are two
configurations of strong ordering possible giving the same result, so
we collect them together:
\begin{align}
\A_{\Box}\C_\Box&= 2 \A^{(0)} \, \int \frac{d^d k}{(2 \pi)^d} \bar{f}(k)
\Theta(Q^2-\ka^2)\Theta(\ka^2-Q_0^2)\C_\Box,\\ 
&\bar{f}(k) = i g^2_s \mu^{2 \ep} \;\frac{p_1\cdot
p_2}{(p_1k-i\ep)(p_2k+i\ep)(k^2+i\ep)};\label{def fb}\\
\A_{\Cross}\C_{\Cross}&= 2 \A^{(0)} \,\int \frac{d^d k}{(2 \pi)^d}
f(k) \Theta(Q^2-\ka^2)\Theta(\ka^2-Q_0^2)\C_{\Cross},\\
&f(k) = - i g^2_s \mu^{2 \ep} \;\frac{p_1\cdot
p_2}{(p_1k-i\ep)(p_2k-i\ep)(k^2+i\ep)}.\label{def f}
\end{align}
We have again extracted the colour factors explicitly, $\C_\Box$
and $\C_\Cross$.  $\ka$ is the usual transverse part of $k$ ($\ka^2>0$). The
upper  limit on $\ka^2$ as imposed by the $\Theta$ function is a
consequence of the strong ordering condition. Simple colour algebra gives
\begin{eqnarray}
  \C_\Box   &=& \;\;\,\phantom{\frac{N_c}2}-\frac1{N_c}\;\;\,\C_8
                +\frac{N_c^2-1}{4N_c^2}\C_1, \\
  \C_\Cross &=& \left(\frac{N_c}2-\frac1{N_c}\right)\C_8
                +\frac{N_c^2-1}{4N_c^2}\C_1.
\end{eqnarray}
Putting together the terms and performing some simplification, we
arrive at
\begin{eqnarray}
  \A^{(1)}\C &\equiv& \A_\Box\C_\Box+i\A_\Cross\C_\Cross\\
  &=& \A^{(0)}\left(\frac{N_c^2-1}{4N_c^2}\G\,\C_1 +
  \left(\frac{N_c}2\F-\frac1{N_c}\G\right)\C_8 \right), 
\end{eqnarray}
with
\begin{eqnarray}
  \F &=& 2 \int\frac{\d^dk}{(2\pi)^d} f(k) \;\Theta(Q^2-\ka^2)\Theta(\ka^2-Q_0^2), \\
  \G &=& 2 \int\frac{\d^dk}{(2\pi)^d} [ f(k) + \bar{f}(k) ]\;
  \Theta(Q^2-\ka^2)\Theta(\ka^2-Q_0^2)\label{def F}
\end{eqnarray}
We introduce Sudakov variables and perform the integration with respect to two
of them. The angular integration in the transverse momentum integral
is trivial and as the result for $\F$ and $\G$ we obtain ($d=4-2\ep$):
\begin{align}
  \F &= -\frac{2\as}{\pi}\;\frac{(4\pi\mu^2)^\ep}{\Gamma(1-\ep)} \,Y\,
  \int^Q_{Q_0}\frac{\d k}{k^{1+2\ep}}\label{res F}\\
  \G &= -\frac{2\as}{\pi}\;\frac{(4\pi\mu^2)^\ep}{\Gamma(1-\ep)} \,(i\pi)\,
  \int^Q_{Q_0}\frac{\d k}{k^{1+2\ep}}\label{res G}
\end{align}
Note that the real parts from $f$ and $\bar{f}$ in $\G$ (\ref{def F})
and hence the terms proportional to $Y$ have cancelled. 
\subsection{The two loop amplitude}
\begin{center}
\EPSFIGURE[h]{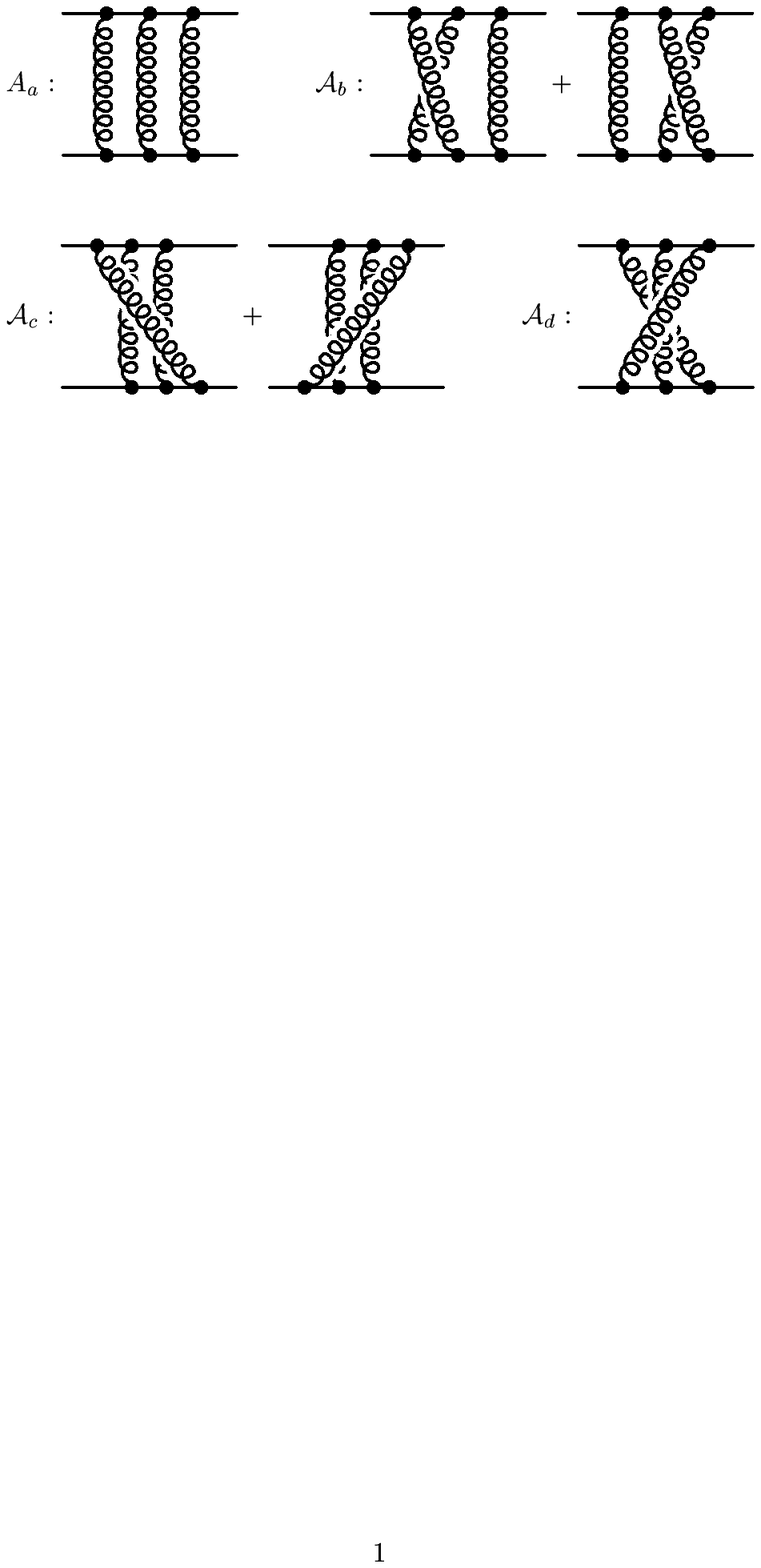}{The 2-loop amplitude\label{fig 2-loop}}
\end{center}
According to the colour structure there are four groups of Feynman
diagrams (Fig.\ref{fig 2-loop}). For each group there are four 
configurations of strong ordering giving the same result.
In terms of the functions $f$ and $\bar{f}$, (\ref{def f},\ref{def
 fb}) the amplitudes read:
\begin{align*}
\A^{(2)}_x\C_x&=4\A^{(0)}\int \frac{d^d k_1}{(2 \pi)^d }\int \frac{d^d
k_2}{(2 \pi)^d } \;a_x(k_1, k_2) \;\Theta(Q^2-\ka_1^2)\Theta(\ka_1^2-\ka_2^2)\Theta(\ka_2^2-Q_0^2)\;\C_x\\
&x=a, b, c, d
\end{align*}
with 
\begin{align}
a_a(k_1, k_2) &= \bar{f}(k_1) \bar{f}(k_2)\\
a_b(k_1, k_2) &= f(k_1) \bar{f}(k_2)\\
a_c(k_1, k_2) &= \bar{f}(k_1) f(k_2)\\
a_d(k_1, k_2) &= f(k_1) f(k_2)
\end{align}
 We can
express each amplitude in terms of $\F$ and $\G$ since
the $\Theta$-functions only affect the integration over the  absolute
value of the transverse momentum which we left undone in  $\F$ and
$\G$, (\ref{res F},\ref{res G}). The decomposition of the sum of
the four amplitudes in terms of $\C_8$ and $\C_1$
is straightforward. We finally find for the two-loop
amplitude ($\C$ being the corresponding colour factor):
\begin{align} 
\Re (\A^{(2)}\C) &=  \A^{(0)} \frac{1}{2}\left[ |\G|^2 \frac{N_c^2-1}{4
N_c^3} \C_1 +  \left(\frac{N_c^2}{4} \F^2 
- \frac{N_c^2+3}{4 N_c^2}  |\G|^2 \right) \C_8 \right] \\
\Im (\A^{(2)}\C) &=  \A^{(0)} \frac{1}{2} |\G| \F \left[
\frac{N_c^2-1}{8 N_c} \C_1 - \C_8 \right].
\end{align}
Note that the sub-leading terms in the real part of $\A^{(2)}$
stem from keeping the sub-leading imaginary part of the one loop result. 

\acknowledgments
MHS thanks R.B.Appleby for useful discussions. 
AK would like to thank PPARC for supporting part of this research.

\end{document}